\documentclass[12pt]{article}
 \pdfoutput=1

\usepackage{amsmath,amssymb,amsfonts}
\usepackage{graphicx}
\usepackage{color}
\usepackage{multirow}
\usepackage{hyperref}

\newcommand{\eq}[1]{Eq.~(\ref{#1})}

\newcommand{\barray}{\begin{eqnarray}}
\newcommand{\earray}{\end{eqnarray}}
\newcommand{\nn}{\nonumber \\}

\newcommand{\beq}{\begin{equation}}
\newcommand{\eeq}{\end{equation}}
\newcommand{\ba}{\begin{array}}
\newcommand{\ea}{\end{array}}
\newcommand{\bea}{\begin{eqnarray}}
\newcommand{\eea}{\end{eqnarray} }
\newcommand{\be}{\begin{eqnarray}}
\newcommand{\ee}{\end{eqnarray} }
\newcommand{\bal}{\begin{align}}
\newcommand{\eal}{\end{align}}
\newcommand{\bi}{\begin{itemize}}
\newcommand{\ei}{\end{itemize}}
\newcommand{\ben}{\begin{enumerate}}
\newcommand{\een}{\end{enumerate}}
\newcommand{\bc}{\begin{center}}
\newcommand{\ec}{\end{center}}
\newcommand{\bt}{\begin{table}}
\newcommand{\et}{\end{table}}
\newcommand{\btb}{\begin{tabular}}
\newcommand{\etb}{\end{tabular}}

\def\cl{{\mathcal L}}

\newcommand{\cM}{{\mathcal M}}

\def\pa{\partial}

\newcommand{\eps}{\epsilon}

\numberwithin{equation}{section}

 \usepackage{hyperref}
\definecolor{rossos}{rgb}{0.7,0,0.3}
\definecolor{bluscuro}{rgb}{0.15, 0.2, 0.9}
\definecolor{verdes}{rgb}{0.1, 0.5, 0.1}
 \hypersetup{colorlinks, citecolor=bluscuro, linkcolor=bluscuro, urlcolor=bluscuro}

\begin{document}

\begin{titlepage}
\vspace{-1cm}
\begin{flushright}
\small
LPT-ORSAY 13-20
\end{flushright}
\vspace{0.2cm}
\begin{center}
{\Large \bf Higgs At Last}
\vspace*{0.2cm}
\end{center}
\vskip0.2cm

\begin{center}
{\bf Adam Falkowski\,$^{a}$, Francesco Riva\,$^{b}$,  Alfredo Urbano\,$^{c}$}

\end{center}
\vskip 8pt

\begin{center}
{\it $^{a}$ Laboratoire de Physique Th\'eorique d'Orsay, UMR8627--CNRS,\\ Universit\'e Paris--Sud, Orsay, France.}\\
{\it {$^b$\,  Institut de Th\'eorie des Ph\'enom\`enes Physiques, EPFL,1015 Lausanne, Switzerland.}}\\
{\it {$^c$\,  SISSA, via Bonomea 265, I-34136 Trieste, Italy.}}\\
\end{center}

\vspace*{0.3cm}

\vglue 0.3truecm

\begin{abstract}
\vskip 3pt \noindent

We update the experimental constraints on the parameters of the Higgs effective Lagrangian. 
We combine the most recent LHC Higgs data in all available search channels with electroweak precision observables from SLC, LEP-1, LEP-2, and the Tevatron. 
Overall, the data are perfectly consistent with the 126 GeV particle being the Standard Model Higgs boson. 
The Higgs coupling to W and Z bosons relative to the Standard Model one is constrained in the range $[0.98,1.08]$ at 95\% confidence level, independently of the values of other Higgs couplings.    
Higher-order Higgs couplings to electroweak gauge bosons are also well constrained by a combination of LHC Higgs data and electroweak precision tests. 

\end{abstract}

\end{titlepage}

\newpage

\section{Introduction}

The existence of a boson with a mass around $m_h=126$~GeV is firmly established \cite{higgs}. 
The focus now is on determining the properties of the new particle, in particular its couplings to the Standard Model (SM) matter.  
If new physics beyond the SM plays a role in electroweak (EW)  symmetry breaking, then Higgs couplings may be modified in an observable way.   
So far, the measured couplings of the new particle are consistent with those of the SM Higgs boson;  
nevertheless,  the current experimental precision leaves ample room for new physics. 

A general framework to study potential deviations of Higgs couplings from the SM is that of an effective theory.
The basic assumption behind this approach is that new degrees of freedom coupled to the Higgs are heavy enough such that their effects can be described by means of local operators involving the SM fields only. 
These operators can be organized into a formal derivative expansion, according to the relevance for Higgs observables: the leading order (LO) operators with no derivatives, the next-to-leading order (NLO) operators suppressed by two derivatives, and so on. 
Previous studies along these lines have shown that the coefficients of the leading operators in this expansion can already be meaningfully constrained  \cite{higgsfits,higgsewfits}.  

The purpose of this paper is two-fold.  
One is to update the constraints on the effective theory parameters using the most recent Higgs data from the LHC  \cite{CMSnew1}-\cite{ATLAS_coup}. The other is to point out that not only the LO but also some NLO  operators in the effective theory can be meaningfully constrained using the existing data. 
This can be achieved by combining the LHC Higgs data and EW precision constraints.  
In this paper we  concentrate on the class of models that are favored by EW precision data and study the constraints from SLC, LEP-1, LEP-2 and $W$ mass measurement on the Higgs couplings, including the very recent update of the fermionic cross section measurements at LEP-2 \cite{Schael:2013ita}.  We will argue that Higgs and electroweak data taken together impose strong bounds on the coefficients of the LO and NLO operators coupling the Higgs boson to the SM gauge bosons.

The paper is organized as follows.  
In Sec.~2  we introduce  a general Higgs low-energy effective Lagrangian, subject only to minimal assumptions about flavor and some EW precision constraints.
In Sec.~3 we further discuss the bounds on the parameters of our Lagrangian from electroweak precision tests; to this end we perform a comprehensive and up-to-date fit to the EW observables. 
In Sec.~4 we summarize the dependence of Higgs physics observables on the parameters of the effective Lagrangian. 
In Sec.~5  we present combined fits to the parameters of our effective Lagrangian from the  latest Higgs data, combined with EW precision measurements. 
Conclusions  follow in Sec.~6, while in Appendix \ref{app:PMS} we discuss the theoretical basis of the effective Lagrangian used in our phenomenological analysis and Appendix \ref{app:PO} gives a summary of the EW oblique parameters.

\section{Effective Theory}
\label{sec:efflag}

A general and convenient framework for interpreting the experimental Higgs data is that of effective field theories.\footnote{
The effective theory approach to Higgs physics has been pursued before in countless papers, both in the pre-Higgs era (see e.g. \cite{Giudice:2007fh,effectiveold}) as  well as in the context of the 126 GeV Higgs at the LHC (see e.g. \cite{effective125});  the discussion of NLO Higgs couplings to EW gauge bosons is however new of this work (see also Ref.~\cite{Masso:2012eq}). } 
The starting assumption is that  there are no new light degrees of freedom,  other than those of the SM, with significant couplings to the Higgs.   
This allows one to write down the Higgs interaction Lagrangian using the SM degrees of freedom only.    
The effects of integrating out  new physics are encoded in deviations of the LO Higgs couplings from their SM values, and in appearance of NLO (non-renormalizable) Higgs interactions with matter.  
The effective  Lagrangian is organized as a double expansion: in powers of the Higgs boson field $h$, and in the number  of derivatives. 
Since only single Higgs interactions are currently probed, here we ignore all terms of order $h^2$ and higher (as we will argue later, the effect of order-$h^2$ terms on the EW precision parameters can also be neglected in our analysis). 
Thus our effective Lagrangian is organized into a derivative expansion  
\beq
{\cal L}_{\rm Higgs} =  {\cal L }_{(0)} + {\cal L}_{(2)} + \dots 
\eeq    
where each term contains a single power of $h$, the LO term has no derivatives acting on the fields, 
the NLO term has two derivative acting on the fields, and so on. 
According to the effective theory philosophy,  the Lagrangian contains an infinite number of terms and couplings, but only a finite number of terms, at lowest orders in  the derivative expansion, is relevant at a given precision level. 
Given the current experimental precision, stopping at the 2-derivative level is fully adequate.   

In complete generality, we don't assume that the scalar $h$ originates, as in the SM,  from a fundamental $SU(2)_L$ doublet scalar field.    
Our description equally applies if $h$ descends from different representations under the electroweak gauge group (singlet, triplets, etc. ), or is not fundamental   (composite Higgs, dilaton, etc.). Nevertheless, with only a slight abuse of language, we shall refer to it as a Higgs boson.

In this paper we assume that:
\bi 
\item Near 126 GeV there is a unique Higgs boson who is  a color-singlet neutral  scalar with positive parity.     
\item This Higgs has naturally no flavor violating interactions with the SM  fermions. 
\item Higgs interactions obey custodial symmetry under which $h$ is a singlet.   
\ei 
With these assumptions, the lowest-order interaction Lagrangian takes the form: 
\beq 
\label{eq:lp2}
 {\cal L }_{(0)} =   {h \over v} \left [ c_{V} \left ( 2 m_W^2 W_\mu^\dagger W^\mu +  m_Z^2  Z_\mu Z^\mu \right ) 
  - c_t \sum_{f = u,c,t} m_f \bar f f -  c_b  \sum_{f = d,s,b} m_f \bar f f -   c_\tau \sum_{f = e,\mu,\tau} m_f \bar f f  \right ]\, .
\eeq 
As a consequence of custodial symmetry, only one parameter $c_V$ controls the LO couplings  to both $W$ and $Z$ bosons;
relaxing this  would lead to quadratically divergent corrections to the $T$ parameter, and thus any departure from custodial symmetry is severely constrained at the level of $1\%$, barring large fine-tuned cancellations.  
Furthermore, while we allow the couplings to up-type quarks, down-type quarks, and leptons to be independent, we assume that within each of these classes the coupling ratios are equal to  the fermion mass ratio.  Relaxing this would generically lead to flavor-changing Higgs interactions in the mass eigenstate basis, which are very constrained unless  some underlying flavor principle is at work to suppress these dangerous effects.

At the NLO in the derivative expansion  we include
\beq 
\label{eq:lp4}
 {\cal L }_{(2)} =  - {h \over  4 v}   \left [ 2 c_{WW} W_{\mu\nu}^\dagger W^{\mu\nu} + c_{ZZ} Z_{\mu \nu}Z^{\mu \nu} +  2 c_{Z \gamma} A_{\mu\nu} Z^{\mu\nu} +  c_{\gamma \gamma}   A_{\mu\nu} A^{\mu\nu}  -  c_{gg}  G_{\mu \nu}^a G^{a,\mu\nu}   \right ],   
 \eeq 
where custodial symmetry imposes two further restrictions on the couplings (see Appendix~\ref{app:PMS}),
 \beq\label{eq:conditions}
 c_{WW}  =  c_{\gamma\gamma} + {g_L \over g_Y} c_{Z\gamma}\,,  \qquad 
 c_{ZZ} = c_{\gamma\gamma} +  {g_L^2 - g_Y^2 \over  g_L g_Y} c_{Z\gamma}\,,
 \eeq
 with $g_L$, $g_Y$  the $SU(2)_L \times U(1)_Y$ gauge couplings in the SM. 
Unlike in Eq.~\eqref{eq:lp2}, the terms in Eq.~\eqref{eq:lp4} are not the most general interactions terms at the 2-derivative level.
Indeed, terms of the form $h Z_\mu \pa_\nu V^{\mu \nu}$ or $h\, \pa_\nu Z_\mu\pa^\mu Z^\nu$ are omitted because they lead, even after imposing custodial symmetry, to quadratic divergences in the $S$ parameter.  
 A more extensive discussion of all possible interactions, their expected size and phenomenological restrictions is given in Appendix~\ref{app:PMS}; notice in particular that in the majority of theories the coefficients of \eq{eq:lp4} are expected to arise only at the loop level.   
Furthermore, the NLO Higgs interactions with  fermions are neglected, because the current data barely constrains the LO fermionic interactions.
The SM corresponds to the limit where $c_{gg} = c_{\gamma\gamma} = c_{Z \gamma} = 0$ in \eq{eq:lp4}, while 
 $c_V = c_t = c_b = c_\tau =1$ in \eq{eq:lp2}.

Thus, a combination of reasonable assumptions, effective theory arguments, and phenomenological constraints leads us to the effective Higgs interaction Lagrangian with only 7 free parameters:
\beq
c_V, \quad c_t, \quad c_b, \quad c_\tau, \quad c_{gg}, \quad c_{\gamma \gamma}, \quad c_{Z \gamma}. 
\eeq
 In the remainder of this paper we discuss  how data constrains these parameters.  

\section{Electroweak Precision Tests}
\label{sec:ewpt}

It is well known that  electroweak precision observables  are sensitive to the mass and couplings of the Higgs boson. 
At the technical level, the largest effect enters via 1-loop contributions to the 2-point functions of the electroweak gauge bosons, the so-called oblique corrections. 
Once the single Higgs couplings deviate from the SM, the sum of the Higgs and pure gauge contributions to precision observables becomes divergent.
Indeed, as discussed in Appendix~\ref{app:PMS},  custodial breaking parameters such as $c_V^z-c_V^w$ (and at NLO couplings such as $hZ_\mu\partial_\nu V^{\mu\nu}$) would give too large a contribution to the EW precision parameters and have been neglected in our Lagrangian \eq{eq:lp2} and \eq{eq:lp4}.  Furthermore we assume that  other contributions to the electroweak precision parameters, such as tree-level effects or loop effects involving only gauge bosons, are negligible (the presence of such contributions would weaken the connection between EW precision tests and LHC data that we assume here). Therefore, in our Lagrangian, the divergences are only logarithmic, thanks  to custodial symmetry and forbidding this class of  operators. 
As for the possible $h^2$ interactions with the gauge bosons (which we ignore in this paper),  we argue here that in a natural situation they  do not play a role.   
Effective operators involving two Higgs bosons of the type $h^2 V_{\mu}V^{\mu}$, or $h^2 V_{\mu\nu}V^{\mu\nu}$ give a contribution to the gauge bosons two-point functions that is always proportional to the combination $\Lambda^2 - m_h^2\log(\Lambda^2/m_h^2)$. Therefore, imposing the  cancellation of quadratic divergences always cancels the logarithmic terms as well.  Effective operators of the type  $ h(\partial_{\mu}h)V^{\mu}$  do not give any contribution at all. 
Thus, barring an unnatural situation with  quadratically divergent contributions to precision observables,  it is enough to consider the electroweak constraint on the parameters $c_V$, $c_{\gamma \gamma}$, $c_{Z \gamma}$ in our effective Lagrangian. 
 
It is instructive to view the 1-loop Higgs contributions to the oblique parameters $S$, $T$, $W$, $Y$ defined in \cite{Barbieri:2004qk}\footnote{%
See Appendix~\ref{app:PO} for the definition and a discussion of these parameters.  The more familiar $S,T,U$ parametrization is not adequate in our case, since $U$ is not logarithmically enhanced, unlike $W$ and $Y$.}.  
\bea
\label{eq:styw_3par}
 \alpha T &\approx& {3 g_Y^2 \over 32 \pi^2 } \left (c_V^2 - 1 \right )  \log (\Lambda/m_Z) \, ,
 \nn 
 \alpha S &\approx&   {g_L g_Y  \over 48  \pi^2 (g_L^2 + g_Y^2)}
 \left \{ 2 g_L g_Y  \left (1 - c_V^2 \right )  +  6 c_V \left [ 2 g_L g_Y c_{\gamma \gamma}  + c_{Z\gamma} (g_L^2 -  g_Y^2) \right ]  
 \right . \nn & & \left. 
 + 3 \left [ g_L g_Y (c_{Z \gamma}^2  -c_{\gamma \gamma}^2 ) - (g_L^2 - g_Y^2)c_{\gamma \gamma}  c_{Z \gamma} \right ]  
\right \}   \log (\Lambda/m_Z)\, ,
  \nn 
\alpha W  &\approx&  {g_L^2 \over 192 \pi^2} \left ( c_{\gamma \gamma} + {g_L \over g_Y} c_{Z \gamma} \right )^2  \log (\Lambda/m_Z)\, ,
  \nn 
\alpha Y  &\approx&    {g_L^2 \over 192 \pi^2} \left ( c_{\gamma \gamma} -  {g_Y \over g_L} c_{Z \gamma} \right )^2  \log (\Lambda/m_Z)\, ,
\eea 
 where we  have omitted finite contributions in the leading-log approximation (see Ref.~\cite{Pich:2012dv} where these contributions, including large tree-level effects, have been computed in an explicit model using Weinberg sum rules).
These oblique parameters are constrained by electroweak precision data to be small at the level of $0.1-0.2$. 
For the cut-off scale on the order of a TeV  this translates into similar constraints on the effective theory parameters.  
Since $S$, $T$, $W$, $Y$ are separately constrained,  Eq.~\eqref{eq:styw_3par} makes it clear that one can separately constrain each of the three couplings $c_V$, $c_{\gamma \gamma}$, and $c_{Z \gamma}$. 

\begin{figure} 
\bc
\includegraphics[width=0.5\textwidth]{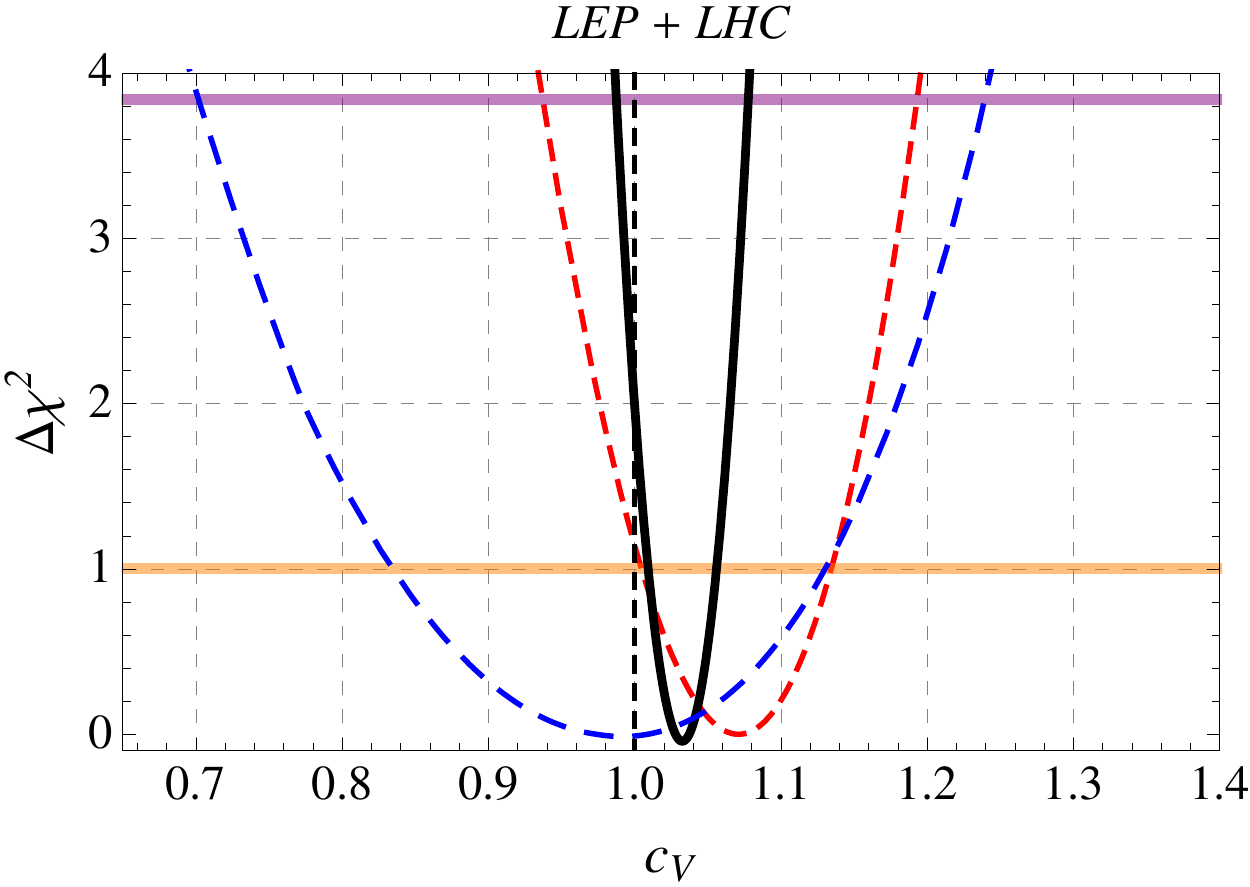}
\includegraphics[width=0.35\textwidth]{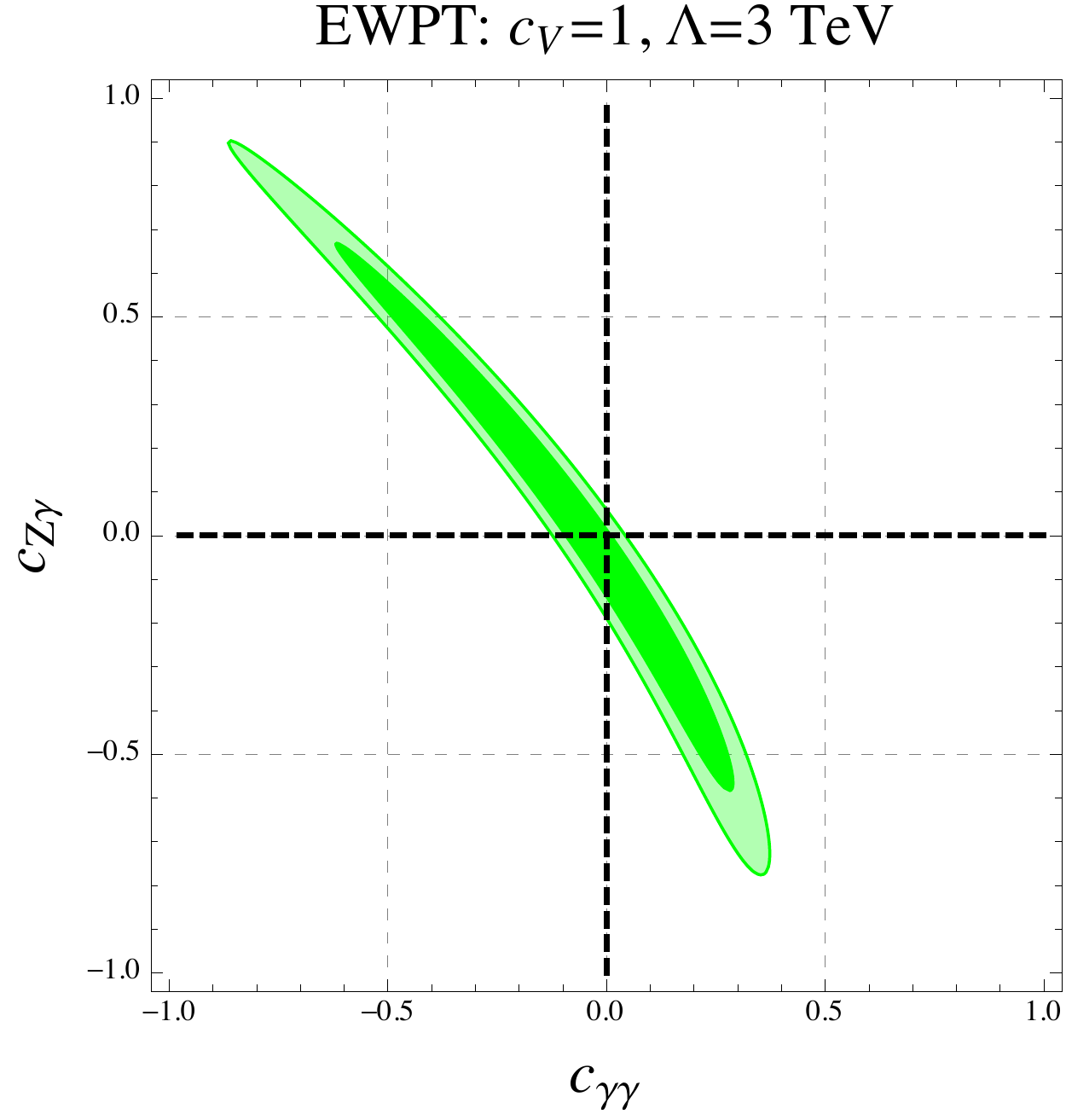}
\ec
\caption{ \footnotesize
Left:  $\chi^2-\chi^2_{\rm min}$ as a function of  $c_V$ for a fit to the Higgs (dashed blue), EW (dashed red), and combined (solid black) data,  after marginalizing over the remaining parameters of the effective theory. 
The orange and purple lines visualize  the 68\% and 95\% CL range of $c_V$.
Right:   Fit of $c_{\gamma \gamma}$ and $c_{Z\gamma}$ to EW data when $c_V$ is fixed to the SM value.   
The $68\%$ (darker green) and $95\%$ CL (lighter green) allowed regions are displayed.  
In both plots $\Lambda = 3$~TeV. 
  }
\label{fig:fit_EWPT}
\end{figure}

In Fig.~\ref{fig:fit_EWPT} we show two examples of  electroweak precision constraints on the parameters of the effective Lagrangian.\footnote{See also \cite{higgsewfits,Masso:2012eq} for previous studies of the impact of electroweak precision tests in the context of the 126 GeV Higgs.}
We use the data from SLC, LEP-1, LEP-2 and W mass measurements summarized in Table~\ref{default}. 
For the cutoff scale $\Lambda = 3$~TeV we find 
\beq
c_V = 1.08 \pm  0.07, 
\qquad 
c_{\gamma \gamma} =  0.10 \pm 0.04, 
\qquad 
c_{Z \gamma} = -0.04 \pm 0.06. 
\eeq 
The limit on the LO coupling $c_V$ are driven by the limits on the T-parameter,  and the $95\%$ CL allowed range is $c_V \in [0.95,1.21]$.
In other words, the couplings of the 125 GeV particle  to the W and Z boson lie within 20\% of those of the SM Higgs, independently of possible inclusion of higher order Higgs couplings, over which we have marginalized.     
These limits are logarithmically sensitive to the cutoff:  for the $95\%$ CL range we obtain   $c_V \in [0.92, 1.30]$ for $\Lambda=1$~TeV, and  $c_V \in [0.96, 1.16]$ for $\Lambda=10$~TeV.
The NLO couplings $c_{\gamma \gamma}$ and $c_{Z \gamma}$ are generically constrained at the level of $0.1$ as well, but  for $c_{\gamma \gamma} \approx - c_{Z \gamma}$ these two coupling are allowed to take ${\cal O}(1)$ values.   
We will see in the following that LHC Higgs data impose a much stronger bound on $c_{\gamma \gamma}$, while the electroweak constraints on  $c_{Z \gamma}$ are competitive to those from the LHC. 

\begin{table}
\begin{center}
\begin{tabular}{|c|c|c|c|}
\hline
{\color{bluscuro}{Observable}} & {\color{bluscuro}{Experimental value}}    &  {\color{bluscuro}{SM prediction}} & {\color{bluscuro}{Description}}    \\  \hline\hline
$m_{W}$ [GeV]  & $80.385 \pm 0.015$    &  $80.3602$ & pole $W$ mass \\ 
\hline \hline 
$\sin^2\theta^{l}_{eff}$  & $0.23116\pm 0.00023$  & $0.231516$ & effective weak mixing angle at $Z$ peak \\ \hline
$\Gamma_Z$ [GeV]  & $2.4952\pm 0.0023$ & $2.49531$   & total $Z$ width \\  \hline
$\sigma_{had}$ [nb]  & $41.540\pm 0.037$ & $41.4778$  & total hadronic cross-section at $Z$ peak\\  \hline 
 $R_{l}$  & $20.767\pm 0.025$ & $20.7408$ & $\Gamma(Z\to q \bar q)/\Gamma(Z\to l^+ l^-)$ at $Z$ peak 
   \\  \hline
 $A_{l}^{\rm FB}$ & $0.0171\pm 0.0010$ & $0.0162$ & $ l^+ l^-$ forward-backward asym. at $Z$ peak  \\  \hline
$R_b$ & $0.21629\pm0.00066$ & $0.21474$   & $\Gamma(Z \to b\bar{b})/\Gamma(Z\to q \bar q)$  at $Z$ peak  \\  \hline
$R_c$ & $0.1721\pm0.0030$  & $0.1724$ & $\Gamma(Z \to c\bar{c})/\Gamma(Z\to  q \bar q )$ at $Z$ peak  \\  \hline
$A_{b}^{\rm FB}$ & $0.0992\pm 0.0016$ & $0.1032$ & $b\bar{b}$ forward-backward asym. at $Z$ peak  \\  \hline
$A_{c}^{\rm FB}$ & $0.0707\pm 0.0035$  & $0.073$ &$c\bar{c}$ forward-backward asym. at $Z$ peak  \\  \hline
$A_b$ & $0.923\pm 0.020$ & $0.935$ & $b\bar{b}$ polarization asymmetry at $Z$ peak  \\  \hline
$A_c$ & $0.670 \pm 0.027$ & $0.667$  & $c\bar{c}$ polarization asymmetry at $Z$ peak  \\  \hline \hline
$\sigma(\mu^+\mu^-)$  & Table 3.4 Ref.\cite{Schael:2013ita}  &   \cite{Arbuzov:2005ma}  & total $\mu^+\mu^-$ cross-section  at LEP-2  \\ \hline
$\sigma(\tau^+\tau^-)$   &  Table 3.4 Ref.\cite{Schael:2013ita} &  \cite{Arbuzov:2005ma}  & total $ \tau^+\tau^-$ cross-section at LEP-2  \\ \hline
$\sigma(q\bar{q})$  & Table 3.4 Ref.\cite{Schael:2013ita} & \cite{Arbuzov:2005ma}  &  total hadronic cross-section at LEP-2 \\ \hline
$A_{\rm FB}(\mu^+\mu^-)$ & Table 3.4 Ref.\cite{Schael:2013ita}  & \cite{Arbuzov:2005ma}  & $\mu^+\mu^-$ forward-backward asym. at LEP-2   \\ \hline
$A_{\rm FB}(\tau^+\tau^-)$ & Table 3.4 Ref.\cite{Schael:2013ita} &  \cite{Arbuzov:2005ma} & $\tau^+\tau^-$ forward-backward asym. at LEP-2   \\ \hline
$R_b$  & Table 8.9 Ref.\cite{LEP:2003aa}  &  \cite{Arbuzov:2005ma}  & $\sigma(e^+e^-\to b\bar{b})/\sigma(e^+ e^-\to  q \bar q)$ at LEP-2   \\ \hline
$R_c$  & Table 8.10 Ref.\cite{LEP:2003aa}  & \cite{Arbuzov:2005ma}  & $\sigma(e^+e^-\to c\bar{c})/\sigma(e^+ e^-\to  q \bar q)$ at LEP-2   \\ \hline
 $A_{\rm FB}^{b\bar{b}}$ & Table 8.9 Ref.\cite{LEP:2003aa} &  \cite{Arbuzov:2005ma} & $b\bar{b}$ forward-backward asym. at LEP-2   \\ \hline
 $A_{\rm FB}^{c\bar{c}}$ & Table 8.10 Ref.\cite{LEP:2003aa} &  \cite{Arbuzov:2005ma}  & $c\bar{c}$ forward-backward asym. at LEP-2   \\ \hline
${d\sigma(e^+ e^-\to e^+e^-) \over d\cos\theta}$ & Tables 3.8/9 Ref.\cite{Schael:2013ita} &  \cite{Jadach:1995nk} & differential  $e^+e^-$ cross-section at LEP-2  \\ \hline\hline
\end{tabular}
\end{center}
\caption{ \footnotesize
EW precision data included in our fit. 
For the experimental values and errors we used  Ref.~\cite{Group:2012gb} for the W mass,   Ref.~\cite{ALEPH:2005ab} for the  Z-pole observables from SLC and LEP-1, 
Ref.~\cite{LEP:2003aa}   for heavy flavor observables in LEP-2,  and Ref.~\cite{Schael:2013ita} for the remaining LEP-2 observables.  To obtain the SM predictions we used Gfitter for the W mass and Z-pole observables \cite{Baak:2012kk},  and ZFITTER \cite{Arbuzov:2005ma}  for LEP-2 observables except for the $e^+e^- \to e^+e^-$ differential cross section for which we used BHWIDE \cite{Jadach:1995nk}.  
The input parameters are $m_Z=91.1875$~GeV, $e=0.3028$, $v=246.221$~GeV, $m_h=126$~GeV, $m_t = 173.2$~GeV.
} 
\label{default}
\end{table}

\section{Higgs Rates}
\label{sec:higgsrates}

In this section we summarize how the LHC Higgs observables depend on the  parameters of our effective Lagrangian. 
As  customary, we present the results in the form of  rates in various channels relative to the SM ones.

\subsection{Decay}

The widths for fermionic decays mediated by the LO couplings  Eq.~\eqref{eq:lp2} are given by  
\beq
{\Gamma_{cc} \over \Gamma_{cc}^{\rm SM}}   \simeq |c_t|^2, \qquad 
{ \Gamma_{bb}\over \Gamma_{bb}^{\rm SM} }  \simeq |c_b|^2, 
\qquad 
{\Gamma_{\tau \tau} \over \Gamma_{\tau \tau}^{\rm SM}}   \simeq  |c_\tau|^2. 
\eeq 
For  decays into gluons, $\gamma \gamma$, and $Z \gamma$ there are two types of contributions that enter the decay amplitude at the same order in the effective theory counting: 
the one-loop contributions proportional to the couplings $c_i$ of the LO Lagrangian \eqref{eq:lp2}, and the tree-level contributions proportional to   the couplings $c_{ij}$ in the NLO Lagrangian \eqref{eq:lp4}.       
For  $m_h = 126$~GeV  one finds: 
\bea
{ \Gamma_{gg} \over  \Gamma_{g g}^{\rm SM}}  &\simeq & {|\hat c_{gg}|^2 \over |\hat c_{gg,\rm SM}|^2},  
\quad \hat c_{gg} = c_{gg} + 10^{-2} \left [1.28\,c_t - (0.07 - 0.1\,i)\,c_b \right ], \quad   |\hat c_{gg,\rm SM}| \simeq 0.012, 
\nn 
{\Gamma_{\gamma \gamma} \over \Gamma_{\gamma \gamma}^{\rm SM}} &\simeq& {|\hat c_{\gamma \gamma}|^2\over |\hat c_{\gamma \gamma,\rm SM}|^2}, 
\quad \hat c_{\gamma \gamma} = c_{\gamma \gamma} + 10^{-2} \left (0.97\,c_V - 0.21\,c_t \right ), \quad   |\hat c_{\gamma \gamma,\rm SM}| \simeq 0.0076,  
\nn 
{\Gamma_{Z \gamma} \over \Gamma_{Z \gamma}^{\rm SM}} &\simeq& {|\hat c_{Z \gamma}|^2\over |\hat c_{Z \gamma,\rm SM}|^2}, 
\quad \hat c_{Z \gamma} = c_{Z \gamma} + 10^{-2} \left (1.49\,c_V - 0.09\,c_t \right ), \quad   |\hat c_{Z \gamma,\rm SM}| \simeq 0.014. 
\eea
Notice that we neglect the NLO couplings  in  loop diagrams, as these are formally  higher-order contributions. 

The decay amplitudes for the processes $h\to 4l$ and $h\to 2l 2 \nu$ via intermediate $ZZ$ and $WW$ pairs (with one or two gauge bosons off-shell)  receives a leading contribution proportional to the LO coupling $c_V$, and a subleading one proportional to $c_{ZZ}$ and $c_{WW}$.  
Integrating the amplitude squared over the 4-body final state phase space we obtain
\bea
{\Gamma_{ZZ^* \to 4l} \over  \Gamma_{ZZ^* \to 4l}^{\rm SM}} &\simeq & 
c_V^2 + 0.36\,c_V c_{Z\gamma} +  0.26\,c_Vc_{\gamma\gamma} + 0.05\,c_{Z \gamma}^2  +     0.07\, c_{Z \gamma} c_{\gamma \gamma} +  0.02\,c_{\gamma\gamma}^2,
\nn 
{\Gamma_{WW^* \to 2l2\nu} \over  \Gamma_{WW^*\to 2l2\nu}^{\rm SM}} &\simeq & 
c_V^2 + 0.73\,c_V c_{Z\gamma}+  0.38\,c_V c_{\gamma\gamma} + 0.19\,c_{Z \gamma}^2   +     0.20\, c_{Z \gamma} c_{\gamma \gamma} +  0.05\,c_{\gamma\gamma}^2.
\eea 
Apparently, the contribution from  the NLO Higgs couplings to these decays is suppressed by the off-shellness of the intermediate gauge boson(s), thus for $c_{\gamma \gamma}$ and $c_{Z \gamma}$ of natural size their effect on the $h\to VV^*$ decay widths is expected to be negligible.\footnote{Moreover, since the experimental cuts are tailored for SM-like Higgs coupling, the efficiency of detecting the leptonic final state may be smaller when higher-order Higgs couplings dominate the amplitude. We don't take into account this effect.}

Finally, we recall that the branching fraction in a given channel is  ${\rm Br}_{ii}= \Gamma_{ii}/\Gamma_{h}$, where the total decay width  is now given by 
\beq
{\Gamma_{h} \over \Gamma_{h,\rm SM}} \simeq 
 {\Gamma_{bb}\over \Gamma_{bb}^{\rm SM} }  {\rm Br}_{bb}^{\rm SM} 
 +   {\Gamma_{cc}\over \Gamma_{cc}^{\rm SM} }  {\rm Br}_{cc}^{\rm SM}
 +  {\Gamma_{\tau \tau}\over \Gamma_{\tau \tau}^{\rm SM} }  {\rm Br}_{\tau \tau}^{\rm SM}  
+  {\Gamma_{WW^*}\over \Gamma_{WW^*}^{\rm SM} }  {\rm Br}_{WW^*}^{\rm SM} 
+  {\Gamma_{ZZ^*}\over \Gamma_{ZZ^*}^{\rm SM} }  {\rm Br}_{ZZ^*}^{\rm SM}.  
\eeq

\subsection{Production Cross Sections}
\label{s.hpcs}

Given  the current experimental data, the important partonic processes for Higgs production, and their relative cross sections in terms of the effective theory parameters are the following. 
\bi
\item Gluon fusion (ggH), $g g \to h $+jets:
\beq
{\sigma_{ggH} \over \sigma_{ggH}^{\rm SM}} \simeq {|\hat c_g|^2 \over |\hat c_{g,\rm SM}|^2 }. 
\eeq 
\item Vector boson fusion (VBF),  $q q \to h qq$+jets: 
\beq
{\sigma_{VBF} \over \sigma_{VBF}^{\rm SM}} \sim c_V^2 + 0.6\,c_V c_{Z \gamma} + 0.3\,c_V c_{\gamma \gamma} 
 + 3.4\,c_{Z \gamma}^2  +  1.5\,c_{\gamma \gamma}^2 +  3.5\,c_{Z \gamma} c_{\gamma \gamma} .
\eeq 
\item Vector boson associated production (VH), $q \bar q \to h V$+jets, where $V=W,Z$,
\bea
{\sigma_{W H} \over \sigma_{WH}^{\rm SM}} & \simeq & c_V^2 -  7.0\,c_V c_{Z \gamma} - 3.6\,c_V c_{\gamma \gamma} 
 + 20.4\,c_{Z \gamma}^2  +  5.5\,c_{\gamma \gamma}^2 +  21.2\,c_{Z \gamma} c_{\gamma \gamma}, 
\nn 
{\sigma_{Z H} \over \sigma_{Z H}^{\rm SM}} & \simeq & c_V^2 -  5.7\,c_V c_{Z \gamma} - 3.4\,  c_V c_{\gamma \gamma} 
 + 14.9\,c_{Z \gamma}^2  +  4.3\,c_{\gamma \gamma}^2 +  15.0\,c_{Z \gamma} c_{\gamma \gamma} .\label{eqVBFrates}
\eea 
\item Top pair associated production (ttH), $g g \to h t \bar t$+jets:
\beq
{\sigma_{ttH} \over \sigma_{ttH}^{\rm SM}} \simeq  c_t^2. 
\eeq
\ei
For  VBF and VH the relative cross sections are given for LHC at $\sqrt{s} =8$~TeV; at 7 TeV the coefficients differ by up to 3\%. 
Furthermore, for the VBF production channel, the relative cross section depends on the cuts on the final states jets; we used $m_{jj} > 400$~ GeV, $|\eta_j| < 4.5$, $\Delta \eta_{jj} > 2.8$. 
Much as for the $h \to VV^*$ decays, large effects from the NLO parameters on the VBF production rate are unlikely. 
For the VH production channel, on the other hand,  ${\cal O}(1)$ corrections  are possible even for $c_V = 1$ and reasonable values of  $c_{\gamma \gamma}$ and $c_{Z \gamma}$. 
Notice also, from \eq{eqVBFrates}, that in the presence of the NLO Higgs interactions, the  $WH$ and $ZH$ production rates relative to the SM can be different without violating custodial symmetry.

\section{Results and discussion}
\label{sec:result}

In this section we present a fully up-to-date fit to the parameters of the Higgs effective Lagrangian. 
We combine the available Higgs results summarized in Table~\ref{tab:higgs}  and EW precision observables from Table~\ref{default}. 
Throughout this section we assume $\Lambda = 3$~TeV to estimate the logarithmically divergent Higgs  contributions to oblique parameters. 

\begin{table}
\begin{center}
\centering
\begin{tabular}{c c}
\begin{tabular}[t]{|c|c|c|c|c|}
\hline
\multicolumn{4}{ |c| }{\textbf{CMS}}   \\ [1 pt] \hline \hline
 & {\color{bluscuro}{Category}} &   
{\color{bluscuro}{$\hat{\mu}$}} & {\color{bluscuro}{Ref.}}   \\ [1 pt] \hline \hline
$\gamma \gamma$ & VBF+VH/ggF  & $0.77^{+0.29}_{-0.26}$& \cite{CMSnew1} \\
\hline
\multirow{3}{*}{$WW$} & 0/1j & $0.73^{+0.22}_{-0.20} $ & \multirow{3}{*}{\cite{CMSnew5}}    \\
& VBF  & $-0.05^{+0.75}_{-0.56}$ & \\
& VH  & $0.51^{+1.26}_{-0.94}$ &  \\ 
\hline
\multirow{2}{*}{$ZZ$} 
& untag. & $ 0.86^{+ 0.32}_{-0.26}$& \multirow{2}{*}{\cite{CMSnew2}} \\
& dijet  & $1.24^{+0.85}_{-0.58}$& \\  \hline
$Z\gamma$ & incl. & $-1.8^{+5.6}_{-5.6}$& \cite{CMSnew6}  \\ \hline
\multirow{3}{*}{$\tau\tau$} & $0/1 j$ & $0.77^{+0.58}_{-0.55}$& \multirow{3}{*}{\cite{CMSnew4}} \\
& VBF   & $
 1.42^{+0.70}_{-0.64}$& \\
& VH & $0.98^{+1.68}_{-1.50}$& \\ \hline
\multirow{4}{*}{$bb$} & $ ZH(l^+l^-)$  & $1.52^{+1.20}_{-1.082}$ & \multirow{3}{*}{\cite{CMSold44}} \\
& $ ZH(\nu\nu)$  & $1.76^{+1.12}_{-1.00}$&  \\
& $ WH$  & $0.64^{+ 0.92}_{-0.88}$& \\
& $ttH$  & $-0.15^{+2.8}_{-2.9}$ & \cite{CMSnew5} \\
\hline
\end{tabular} &
\begin{tabular}[t]{|c|c|c|c|}
\hline
\multicolumn{4}{ |c| }{\textbf{ATLAS}}   \\ [1 pt] \hline \hline
 & {\color{bluscuro}{Category}} &   
{\color{bluscuro}{$\hat{\mu}$}} & {\color{bluscuro}{Ref.}}   \\ [1 pt] \hline \hline
\multirow{14}{*}{$\gamma\gamma$}
&  UnCe, low $p_{Tt}$    & $(0.5^{+1.4}_{-1.4})0.87^{+0.73}_{-0.70}$& \multirow{14}{*}{\cite{ATLASnew12}}  \\
& UnCe, high $p_{Tt}$  &  $(0.2^{+2.0}_{-1.9})0.96^{+1.07}_{-0.95}$&    \\
&    UnRe, low $p_{Tt}$  &  $(2.5^{+1.7}_{-1.7})2.50^{+0.92}_{-0.77}$&\\
 &  UnRe, high $p_{Tt}$     &  $(10.4^{+3.7}_{-3.7})2.69^{+1.35}_{-1.17}$&\\
&CoCe, low $p_{Tt}$      &  $(6.1^{+2.7}_{-2.7})1.39^{+1.01}_{-0.95}$&\\
&CoCe, high $p_{Tt}$  &  $(-4.4^{+1.8}_{-1.8})1.98^{+1.54}_{-1.26}$&\\
&CoRe, low $p_{Tt}$   &  $(2.7^{+2.0}_{-2.0})2.23^{+1.14}_{-1.01}$&\\
&CoRe, high $p_{Tt}$   & $(-1.6^{+2.9}_{-2.9})1.27^{+1.32}_{-1.23}$&\\
&CoTr    & $(0.3^{+3.6}_{-3.6})2.78^{+1.72}_{-1.57}$&\\
&L2j(high mass)      & $2.75^{+1.78}_{-1.38}$&\\
&T2j (high mass)     & $1.61^{+0.83}_{-0.67}$&\\
&2j (low mass)    & $(2.7^{+1.9}_{-1.9}) 0.32^{+1.72}_{-1.44}$&\\
&$E_T^{\rm miss}$ &$2.97^{+2.71}_{-2.15}$&\\
&1l  & $2.69^{+1.97}_{-1.66}$&\\ 
\hline
 $WW$ & VBF+VH/ggF &  $1.35^{+0.57}_{-0.53}$& \cite{ATLASnew30} \\
\hline
$ZZ$ & incl. & $1.35^{+0.39}_{-0.34}$ & \cite{ATLAS_coup} \\
\hline
$Z\gamma$ & incl. &  $2.6^{+6.5}_{-6.5}$ & \cite{ATLASnew9} \\ \hline
$\tau\tau$ & VBF+VH/ggF  & $0.74^{+0.76}_{-0.67}$& \cite{ATLASold160} \\
\hline
$bb$  &VH  & $-0.41^{+1.02}_{-1.04}$& \cite{ATLASold170} \\
\hline
\end{tabular}
\end{tabular}
\end{center}
\caption{  \footnotesize The LHC Higgs data included in our fit \cite{CMSnew1}-\cite{ATLAS_coup}, \cite{CMSold44}-\cite{ATLASold170}.  The rates  are normalized to the SM rate; when data for 7 and 8 TeV are separately provided, we write the former in brackets.  
We also include the latest combined Tevatron measurements: $\hat{\mu}_{\gamma \gamma}= 6.2^{+3.2}_{-3.2}$, $\hat{\mu}_{WW}= 0.9_{-0.8}^{+0.9}$, 
$\hat{\mu}_{bb}^{VH}= 1.62^{+0.77}_{-0.77}$, $\hat{\mu}_{\tau\tau}= 2.1^{+2.2}_{-2.0}$  \cite{Aaltonen:2013kxa}. For the ATLAS $WW$ and $\tau\tau$ and CMS $\gamma \gamma$ channels we include in our fit the two-dimensional likelihood correlations of the signal strengths for the ggF+ttH and VBF+VH production modes. }
\label{tab:higgs}
\end{table}

\subsection{7-parameter fit}


We begin with an unconstrained fit to all seven parameters of the effective Lagrangian.   We find 
\bea & 
c_V =  1.04 \pm 0.03, \quad c_t = 1.1^{+0.9}_{-3.0} \quad c_b = 1.06^{+0.30}_{-0.23}, \quad c_\tau = 1.04 \pm  0.22 
& \nn  & 
c_{gg} = -0.002  \pm 0.036, \quad c_{\gamma \gamma} = 0.0011^{+0.0019}_{-0.0028}, \quad c_{Z \gamma} = 0.000^{+0.019}_{-0.035} \ . 
\eea 
When quoting  $1 \sigma$ errors above we ignored other isolated  minima away from the SM point where a large NLO coupling conspires with the SM loops to produce a small shift of the Higgs observables. 
We find  $\chi_{\rm SM}^2 - \chi_{\rm min}^2 = 4.2$ which means that the SM gives a perfect fit to the Higgs and EW data. 
The previous small discrepancy  due to the enhanced $h\to \gamma \gamma$ rate observed by the ATLAS and CMS goes away after including the latest CMS update in the diphoton channel.  
Remarkably, the current data already puts meaningful limits on {\em all} the parameters.   
The only important  degeneracy is that between $c_{gg}$ and $c_t$: only one linear combination of these two is constrained by stringent limits on the gluon fusion Higgs production, while direct constraints on $c_t$ from searches of the $t \bar t$ associated Higgs production are currently weak. 
Note that the global fit shows a strong preference for  $c_b \neq 0$ even though the $h \to b \bar b$ decay has not been clearly observed by experiment.  
The reason is that $c_b$ dominates the total Higgs width which is indirectly constrained by observing the Higgs signal  in other  decay channels.

The most important result is that  the value of $c_V$ is constrained to be very close to unity, independently of the value of other LO and NLO Higgs couplings.  
 At 95\% CL $c_V$ is constrained in the range $[0.98,1.08]$ by a combination of EW and Higgs data, and in the range $[0.71,1.25]$ by Higgs data alone. 
That means that $h$ is {\em a Higgs boson}, in the sense that it plays a role in EW symmetry breaking. 
More rigorously, $c_V \approx 1$  means that $h$ controls the high-energy behavior of scattering amplitudes of longitudinally polarized W and Z bosons.     
As is clear from  Fig.~\ref{fig:fit_EWPT},   $c_V>0$ with enormous  statistical significance.  

Regarding the NLO couplings, it must be recalled that these coefficient are expected to arise at the loop level and  to be suppressed by the SM gauge couplings. With the appropriate rescalings ($\pi \alpha_s,\pi \alpha$ and $s_w\pi  \alpha$, see Appendix \ref{app:PMS}), we find that the values above correspond to coefficients of a $gg$, $\gamma\gamma$ and $Z\gamma$,  of order $\lesssim 0.3$ $\lesssim 15$ and $\lesssim 6$, respectively. It is interesting to notice, by comparing Figs.~\ref{fig:fit_EWPT} and \ref{fig:fit_loop}, that the strongest constraints on the $c_{\gamma\gamma}$ coupling already come from the LHC rather than from LEP.

\subsection{Loop New Physics}

We move to a more constrained set-up where the LO couplings in the effective Lagrangian  \eq{eq:lp2} are fixed at the SM value, $c_V = c_f=1$, 
whereas the NLO couplings $c_{gg}, c_{\gamma \gamma}, c_{Z \gamma}$ in  \eq{eq:lp4} are allowed to vary. 
This assumption is motivated by new physics models that affect the Higgs production and decay amplitudes only at 1-loop level.
Indeed, new heavy particles with SM quantum numbers that get (at least a part of) their masses from EW breaking and do not mix with the SM matter will affect the effective Lagrangian in this fashion.  
By playing with quantum numbers and masses of the new particles one can easily engineer any set of  values for the NLO couplings.


We find the following best fit values:
\beq
\label{eq:3par}
c_{gg} = - 0.0007 \pm 0.0009,
\quad 
c_{\gamma \gamma} = 0.0009 \pm  0.0008, 
\quad 
c_{Z \gamma} = -0.028^{+0.028}_{-0.016}. 
\eeq 
Also in this more constrained case  the SM point $c_{gg}=c_{\gamma \gamma}=c_{Z \gamma}=0$  has $\chi^2_{\rm SM} - \chi^2_{\rm min} = 2.6$, and thus provides a perfect fit to the data. 
Once the degeneracy with $c_t$ is removed, the constraints on $c_{gg}$ are at the level $10^{-3}$, much as the constraints on $c_{\gamma \gamma}$. 
The remaining NLO coupling $c_{Z \gamma}$ is less constrained, at the level of few$\times10^{-2}$, because, for the time being, the decay $h \to Z \gamma$ is not observed. 
When setting the $1\sigma$ errors in \eq{eq:3par} we ignored other degenerate minima of $\chi^2$  away from the SM-like minimum; we comment on them below.  

\begin{figure}[!h]
    \begin{center}
    \hspace{-1.cm}
      \includegraphics[width=0.47\textwidth]{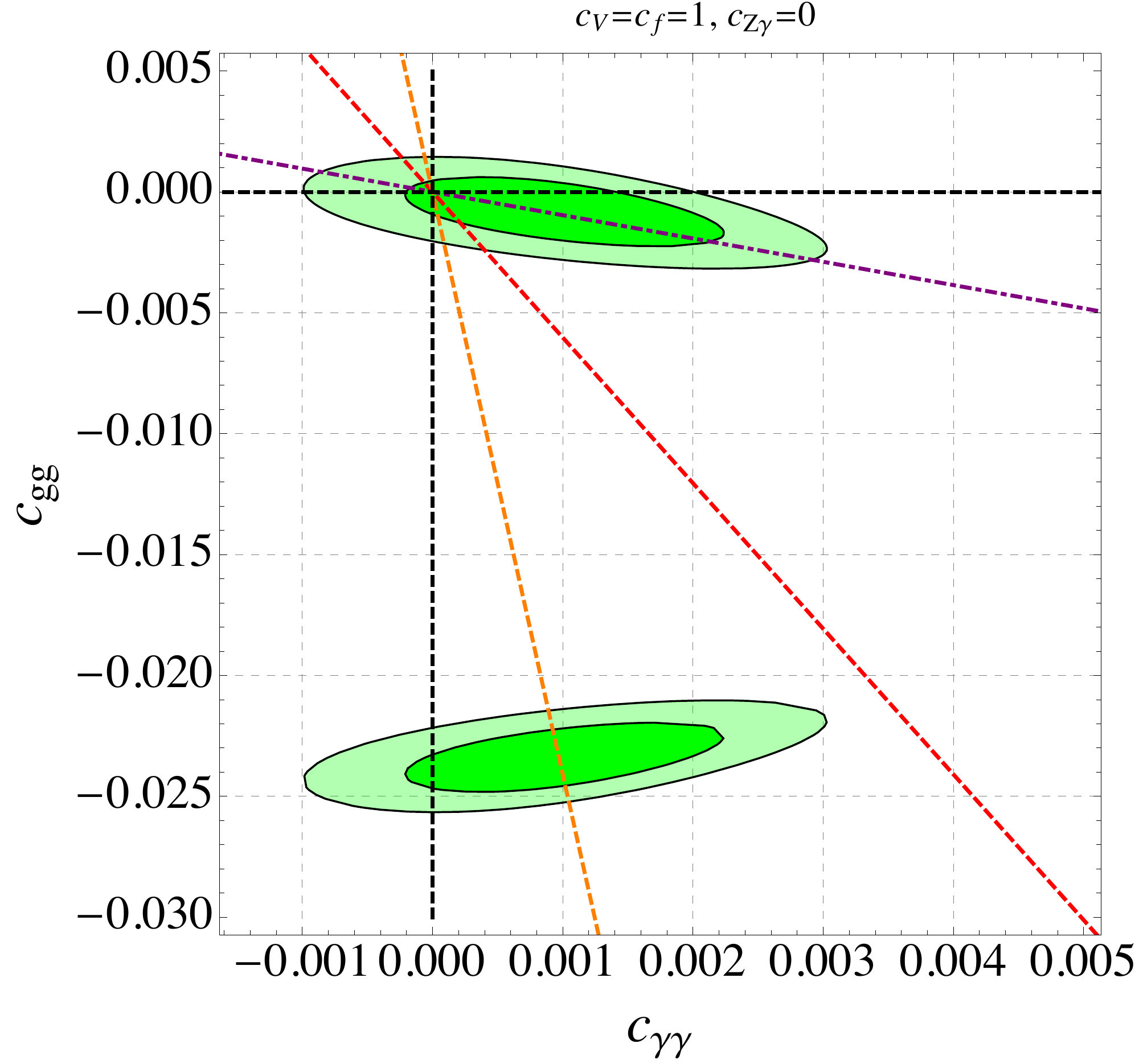}
       \includegraphics[width=0.45\textwidth]{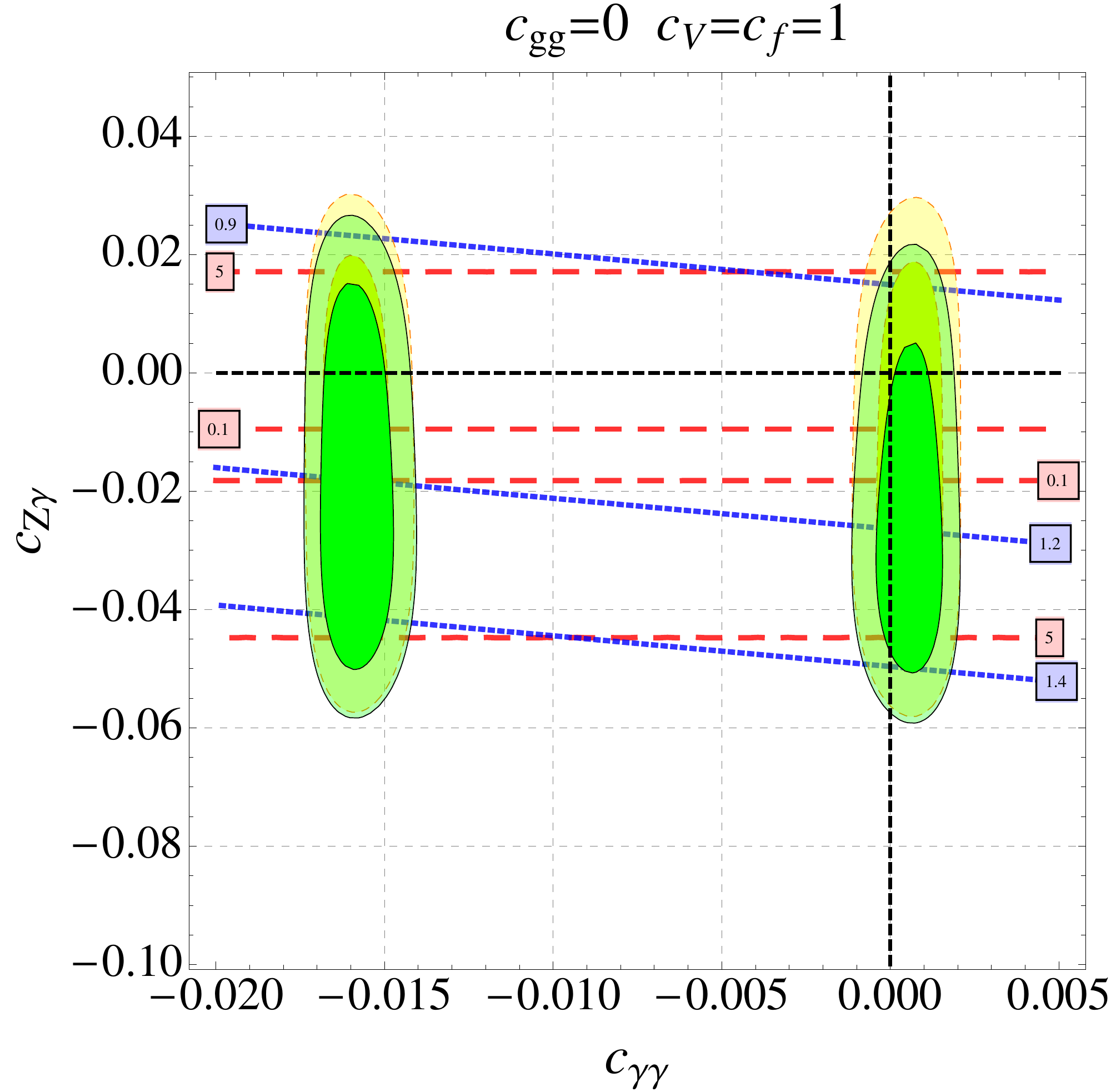}
\vspace*{-2mm}
          \caption{\footnotesize 
          Left:  The 68\% (darker green) and 95\% (lighter green) CL best fit regions in the $c_{\gamma \gamma}$-$c_{g g}$ section of the parameter space.
          The straight lines correspond to a 1-loop contribution from a quark of electric charge $1/3$ (dashed orange), $2/3$ (dotted red), and $5/3$ (dash-dotted purple).   
Right: the same for the $c_{\gamma \gamma}$-$c_{Z \gamma}$ section of parameter space. We also show the contours of $\hat \mu_{Z \gamma}$ (dashed red), and $\sigma_{WH}/\sigma_{WH}^{\rm SM}$ (dotted blue). The yellow regions are fits without the EW data.  }
\label{fig:fit_loop}
\end{center}
\vspace*{-3mm}
\end{figure}

In Fig.~\ref{fig:fit_loop}  we also show two sections of the loop-inspired parameter subspace. 
In the left panel of  Fig.~\ref{fig:fit_loop}, we vary $c_{gg}$ and  $c_{\gamma \gamma}$,   while the least constrained coupling $c_{Z \gamma}$ is set to zero for simplicity. In this case, the EW precision observables have little relevance, as they do not depend on $c_{gg}$ at one loop, while the Higgs constraints on $c_{\gamma \gamma}$ are far more stringent. 
If the NLO couplings are induced by a loop of a single new particle with charge $Q$ and color representation $r$ then  it follows $c_{\gamma \gamma} = - {d(r) Q^2 \over C_2(r)} {\alpha \over \alpha_s}  c_{gg}$. For example, a fermionic or bosonic particle in the same color representation as SM quarks has $C_2(r)=1/2$ and $d(r)=3$. In Fig.~\ref{fig:fit_loop} we display  the trajectories corresponding to the top partner ($Q=2/3$, e.g. stop in SUSY),  bottom partner ($Q=1/3$, e.g. sbottom SUSY), and exotic $Q=5/3$ quarks. The position on the trajectory depends on the mass of the new particle and its couplings to the Higgs, which are arbitrary (up to model-dependent collider and vacuum stability constraints).  
The SM-like best-fit region  the upper one; it contains the SM point $c_{gg} = c_{\gamma \gamma} =0$ within $1 \sigma$ of the best-fit point. 
The other best-fit  region at the bottom  involves fine-tuning in the sense that  a relatively large $c_{gg}$ conspires with the loop contributions proportional to the LO coupling $c_t$ such that the predicted $g g \to h$ rate accidentally falls close to the SM one.     
There are two other  fine-tuned best-fit regions centered at negative $c_{\gamma \gamma}$ that are not displayed in the plot. 
All four best-fit regions have the same $\chi^2$ at the minimum.     

In the right panel of  Fig.~\ref{fig:fit_loop}  we vary $c_{\gamma\gamma}$ and $c_{Z \gamma}$ and set $c_{gg} =0$, which is relevant for integrating out colorless particles. 
The constraints on $c_{Z \gamma}$ from  EW precision observables and from Higgs data are competitive. 
In contrast to the previous case, there are only 2 best-fit regions due the fact there's currently no evidence of the $h \to Z \gamma$ decay.
The $h\to Z \gamma$ rate relative to the SM can be enhanced by a factor of a few, but it could also be suppressed; the Higgs and EW data show no strong preference for either of these possibilities.   
We also show the contours of $WH$ associated production cross section relative to the SM one.  
An enhancement up to 40\% of $WH$ and $ZH$ production due to  NLO Higgs couplings is still perfectly consistent with the  current data.  
Besides, the relative change of the $WH$ and $ZH$ production cross section can be different by up to 5\%, even though custodial symmetry is preserved.  
  
\subsection{Composite Higgs}

Here we present the results  of the 2-parameter fit under the assumption that all the Higgs couplings to fermions take a common value $c_t = c_b = c_\tau \equiv c_f$. 
The other free parameter in this model is the LO  Higgs coupling to $W$ and $Z$ bosons $c_V$. The NLO couplings in the effective Higgs Lagrangian are all set to zero.  
This subspace of the parameter space is inspired by the physics of composite Higgs where a light Higgs boson arises as a pseudo-Goldstone boson of a spontaneously broken approximate global symmetry in a strongly interacting sector \cite{Contino:2010rs}.  
In this class of models, the Higgs coupling to $W$ and $Z$  is suppressed as $c_V = \sqrt{1- \eps^2}$ where $\epsilon = v/f$ and $f$ is called the compositeness scale, or the decay constant of the pseudo-Goldstone boson Higgs. The coupling to fermions also depends on the scale $f$ in a  model dependent way, in particular,  on the representations of the SM fermions under the  global symmetry group of the strongly interacting sector.  For example, for the class of models based on the $SO(5)$ global symmetry broken to $SO(4)$ the fermion couplings take the form \cite{Pomarol:2012qf}:  
\beq
c_f  = {1 + 2m - (1 + 2m + n)\eps^2 \over \sqrt{1- \eps^2}}, 
\eeq 
where $n,m$ are positive integers. The models  with $m \neq 0$ are already excluded by the data, while
$m=0, n=0$ corresponds to the  MCHM4 model and  $m=0, n=1$ to the  MCHM5 model \cite{Agashe:2004rs}. 
Finally, the NLO Higgs couplings of the form $|H|^2 V_{\mu \nu}^2$ that could give rise to non-zero $c_{\gamma \gamma}$, $c_{gg} $ in our effective Lagrangian are expected to be suppressed because they violate the shift symmetry of the pseudo-Goldstone boson Higgs.

\begin{figure} 
\bc
\includegraphics[width=0.34\textwidth]{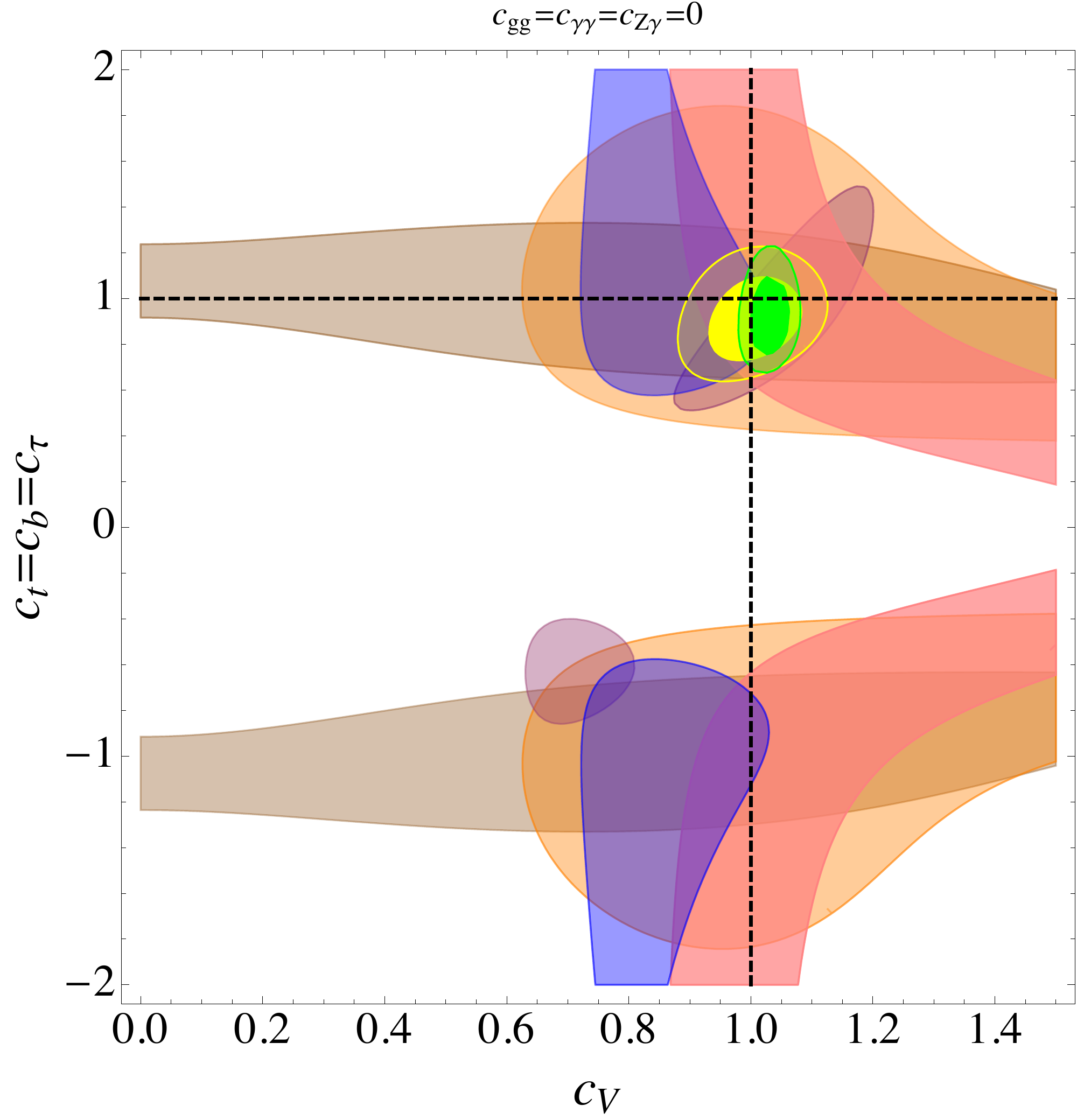} \qquad 
\includegraphics[width=0.5\textwidth]{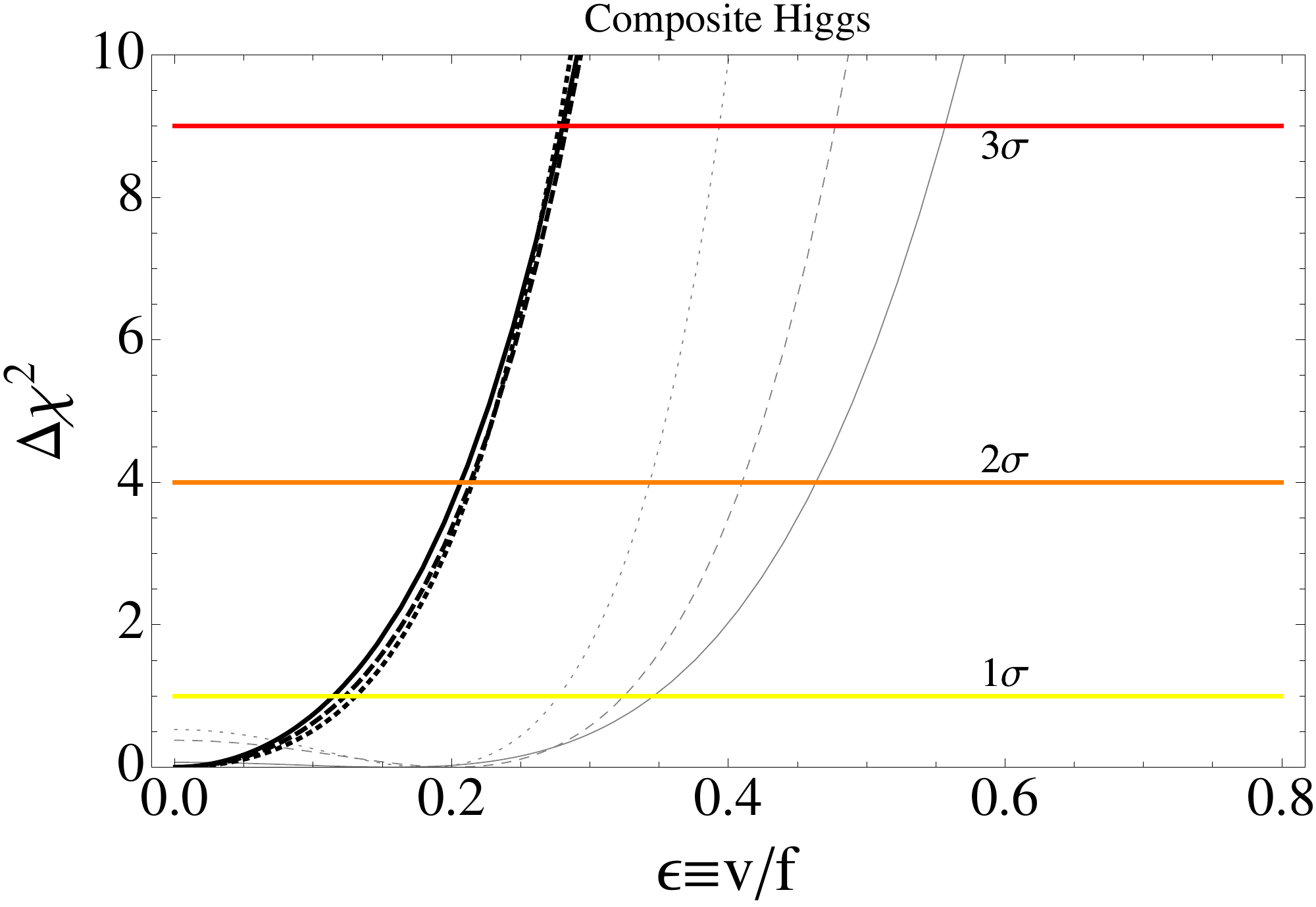}
\ec
\caption{ \footnotesize Left: The 68\% (darker green) and 95\% (lighter green) CL best fit regions in the $c_V$-$c_f$ parameter space. 
The yellow regions are fits without the EW data. 
The color bands are the $1\sigma$ regions preferred by the Higgs data in the $\gamma \gamma$ (purple), $WW$ (blue),  $ZZ$ (red), $\tau \tau$ (brown), and $b b$ (mauve) channels. 
Right: Fit to the parameter $\epsilon= v/f$ in sample composite Higgs models with (black) and without (gray) including EW precision data. The different lines correspond to the $SO(5)/SO(4)$ coset and fermionic representations with $m=0$ and $n=0$ (solid), $n=1$ (dashed) and $n=2$ (dot-dashed)."
  }
\label{fig:fit_cvcf}
\end{figure}

The results of the fit are given in the left panel of  Fig.~\ref{fig:fit_cvcf}. 
The island of good fit for $c_f < 0$ favored by previous Higgs data \cite{higgsfits,higgsewfits}  completely vanishes in the new data. 
The reason is that a large enhancement of the $h \to \gamma \gamma$ rate  is no longer preferred. 
The preference for the SM-like  coupling $c_f \sim 1$ becomes even stronger when EW precision data are included.
This is true in spite of the fact that the EW observables  are not sensitive to $c_f$ at one loop; simply, they prefer $c_V$ somewhat above 1 which is more consistent with the SM island.   In the right panel of  Fig.~\ref{fig:fit_cvcf} we show the bounds on the compositeness scale for a number of composite Higgs models based on the  $SO(5)/SO(4)$ coset.
We find that the strongest bounds still come from EW precision data and, as already pointed out in Ref.~\cite{Barbieri:2007bh}, they push the compositeness scale $f$ at about $1.5$~TeV at 95\% CL, independently of the specific model. Nevertheless, incalculable UV effects could weaken the impact of EW precision data; in this case, and taking into account Higgs data only, the bound on $f$ reduces to the more natural value $f\gtrsim 700$ GeV, with some dependence on the details of the model.

\subsection{2HDM}

Another interesting pattern of couplings is the one where the Higgs couplings to leptons and down-type quarks take a common value $c_d\equiv c_b=c_\tau$ which differs from the coupling to up-type quarks $c_u\equiv c_t$. At the same time,  the  LO coupling to gauge bosons doesn't deviate from the SM, $c_V=1$, and NLO couplings vanish. 
Notice that this plane is insensitive to bounds from EW precision data. This parametrization is inspired by type-II two-Higgs doublets models (2HDM), in particular by minimal supersymmetry. Indeed, in supersymmetric models, assuming heavy superpartners  as presently suggested by experiments, the peculiar two Higgs doublets structure of the scalar sector makes it possible to induce significant modifications of the LO Higgs couplings to fermions, with  very small modifications to $c_V$ and the NLO Higgs coupling. The form of these deviations is well-known, and in the limit where the remaining scalars in the Higgs sector are heavier $m_H\gtrsim v/\sqrt{2}$ and can be integrated away, they reduce to~\cite{SUSYandFits,Gupta},
\begin{equation}\label{ctcb}
c_{b,\tau}\approx1-\delta\tan{\beta } \frac{v^2}{m_H^2}\, ,\quad
c_t\approx1+ \delta\cot{\beta } \frac{v^2}{m_H^2}\, ,
\end{equation}
where $\tan\beta>1$ and $\delta$ depends on the underlying physics that contributes to the Higgs mass (for instance $\delta>0$ in the MSSM and in models with additional D-terms, while $\delta<0$ in the NMSSM) \cite{Gupta}. Corrections to $c_V$ arise at higher order in the $v/m_H$ expansion and are typically very small. 

\begin{figure} 
\bc
\includegraphics[width=0.4\textwidth]{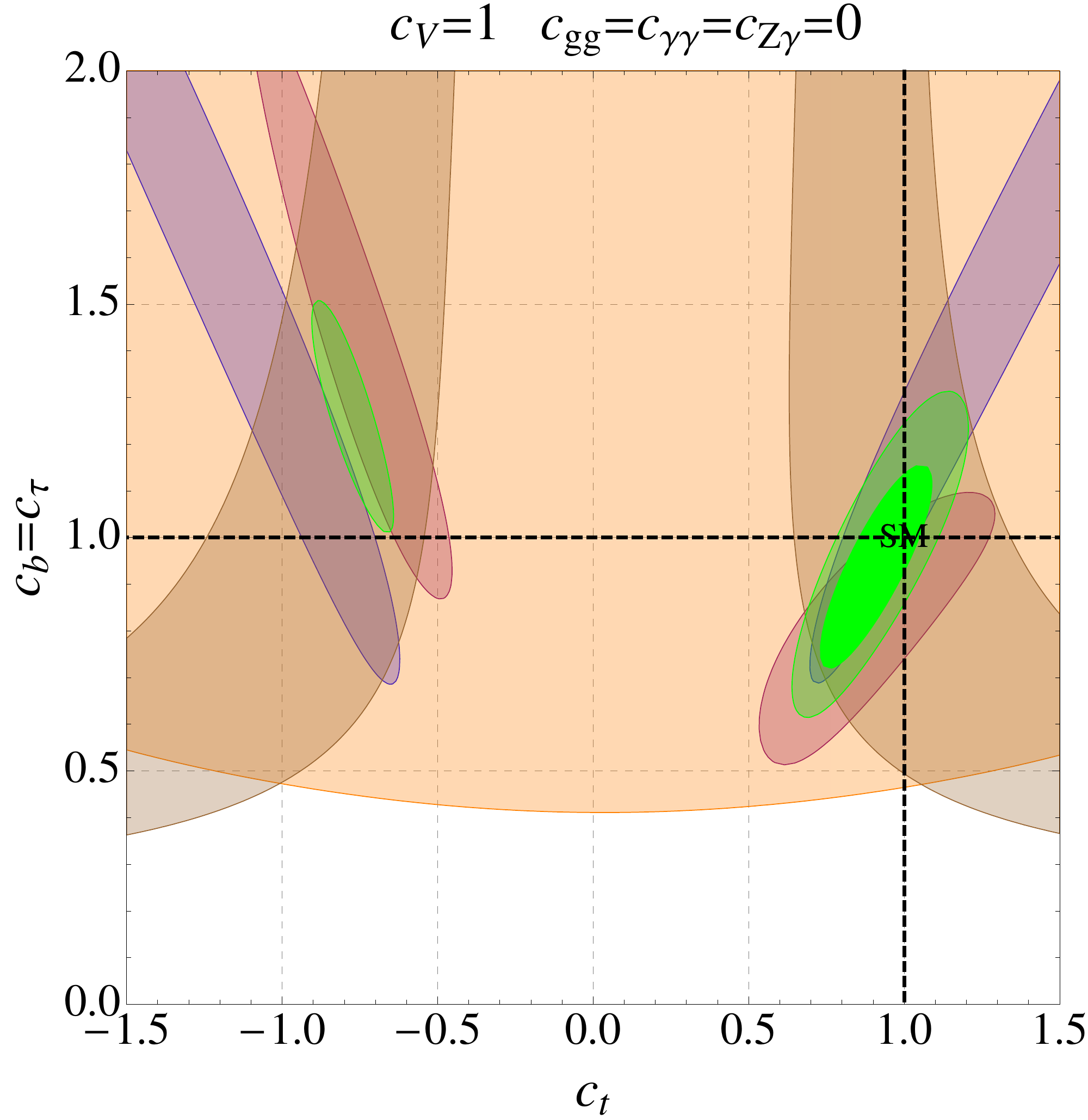} 
\ec
\caption{Left: The 68\% (darker green) and 95\% (lighter green) CL best fit regions in the $c_t$-$c_b=c_\tau$ parameter space. 
The color bands are the $1\sigma$ regions  preferred by the Higgs data in the $\gamma \gamma$ (purple), $VV$ (blue), $\tau \tau$ (brown), and $b b$ (mauve) channels.}
\label{fig:susy}
\end{figure}

We show the fit to $c_b$ and $c_t$ in Fig.~\ref{fig:susy}.  SUSY models imply deviations which lie in the upper-left or lower-right quadrant unless scalar singlets mix with the Higgs.
The  best-fit region at negative $c_t$ preferred in the previous fits \cite{SUSYandFits,Gupta} is now only marginally allowed after including the latest CMS results in the  $\gamma\gamma$ channel \cite{CMSnew1}.

\subsection{Invisible Width} 

In the last of our studies  we are  going beyond our effective Lagrangian and allow Higgs decays to invisible particles.
Searches for such decays are strongly motivated by the so-called Higgs portal models of dark matter (see Ref.~\cite{Frigerio:2012uc} for a natural realization of this scenario).
Furthermore, given the small Higgs width in the SM, $\Gamma_{H,\rm SM} \sim 10^{-5}\,m_h$,  a significant invisible width $\Gamma_{H,\rm inv} \sim \Gamma_{H,\rm SM}$ may easily arise even from small couplings of the Higgs to weakly interacting new physics. In supersymmetric models where the Higgs is the superpartner of a slepton \cite{Riva:2012hz}, there could be invisible decays into neutrinos and gravitinos.

\begin{figure}[!h]
    \begin{center}
    \hspace{-1.cm}
       \includegraphics[width=0.5\textwidth]{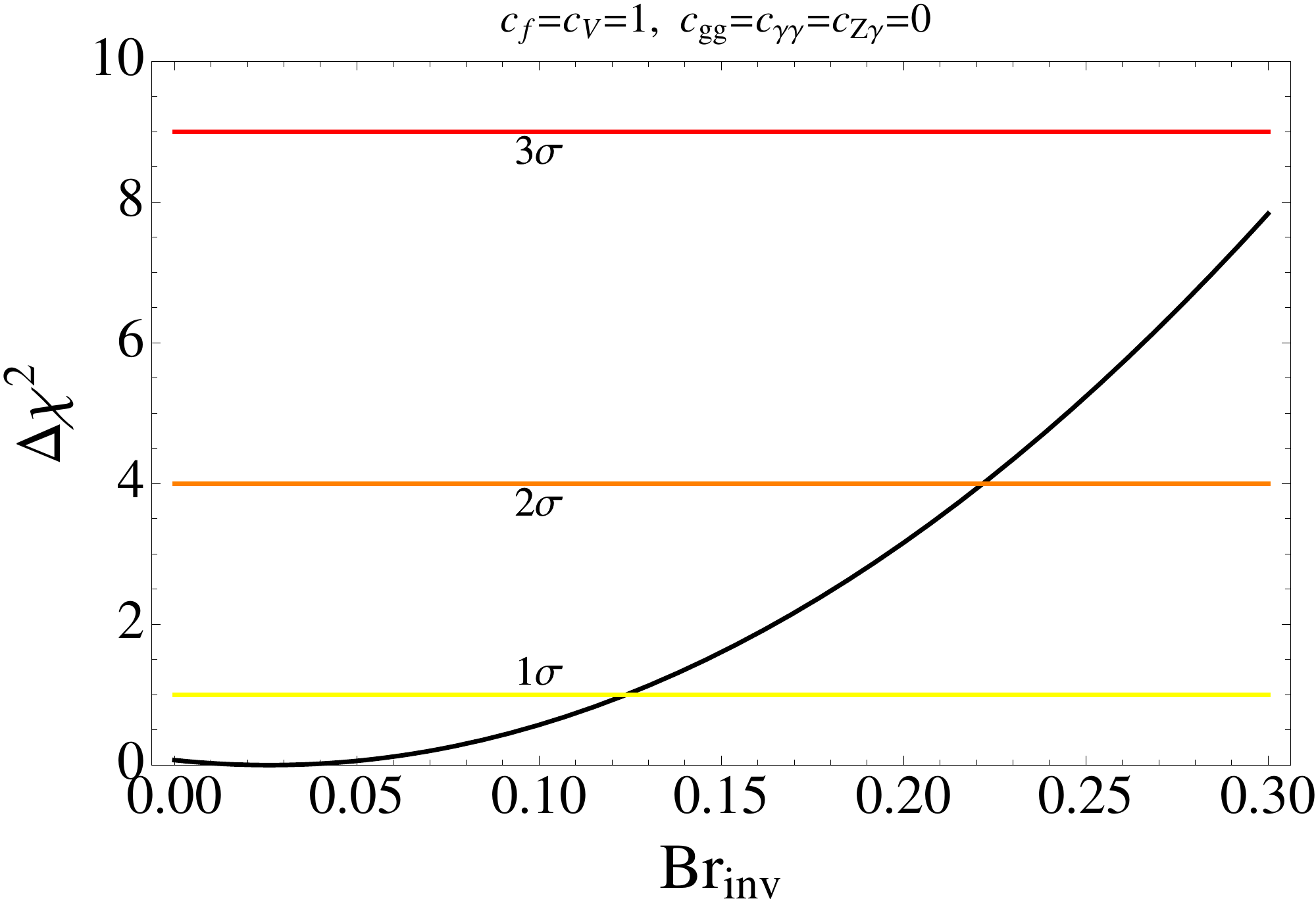}
   \includegraphics[width=0.4\textwidth]{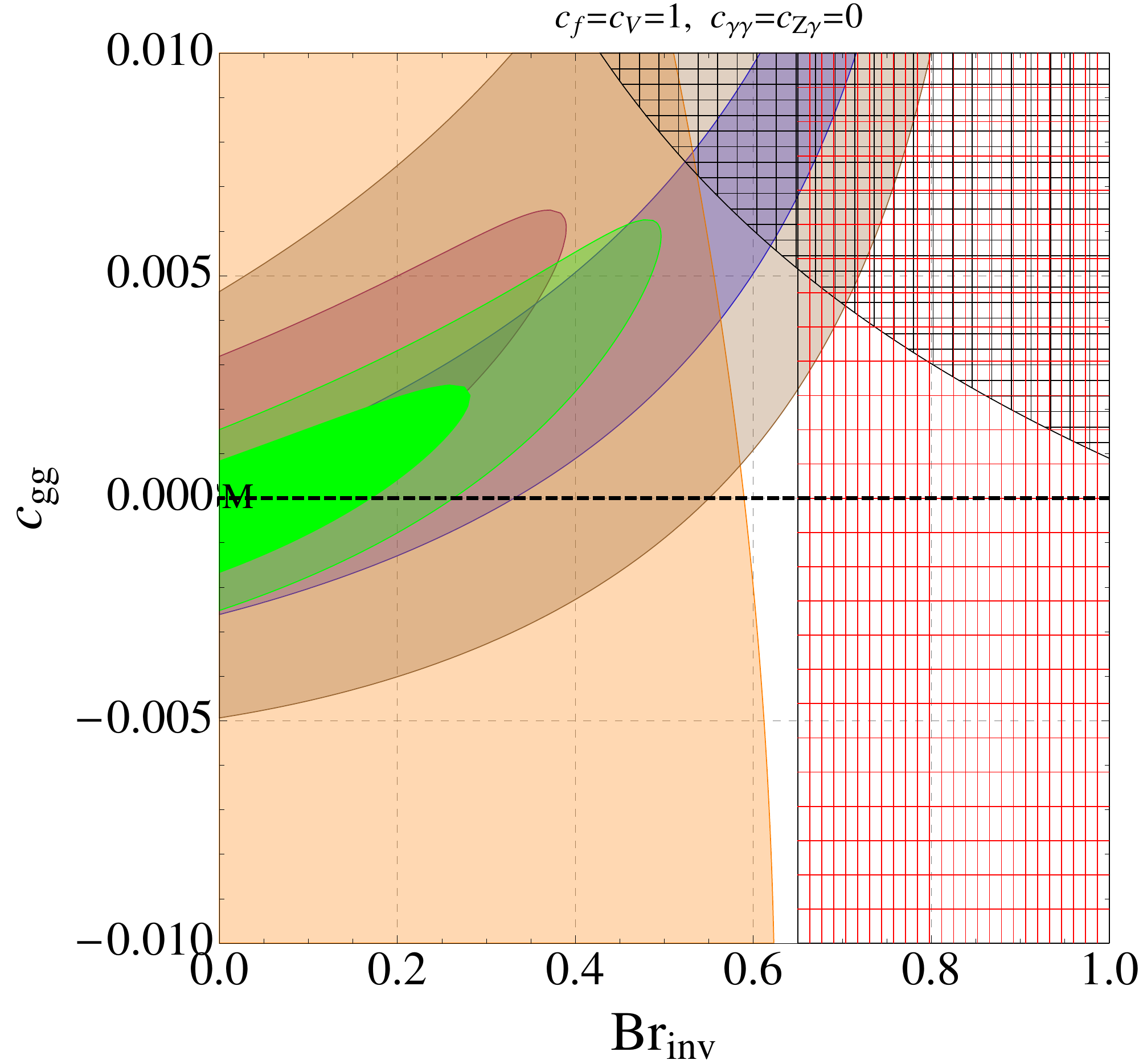}
\vspace*{-2mm}
          \caption{\footnotesize 
          Left:  $\chi^2-\chi^2_{\rm min}$ of the fit for the Higgs with the SM-size couplings to the SM matter and an invisible branching fraction to hidden sector new physics states.  Right: $68\%$ CL (light green) and $95\%$ CL (dark green) best fit regions to the combined LHC Higgs data in a model where the Higgs invisible branching fraction and the NLO coupling to gluons $c_{gg}$  can be simultaneously varied.   
 The color bands are the $1\sigma$ regions preferred by the Higgs data in the $\gamma \gamma$ (purple), $VV$ (blue), $\tau \tau$ (brown), and $b b$ (mauve) channels. 
 The meshed regions are excluded by direct probes of the invisible Higgs width: the ATLAS $Z+h \to {\rm inv.}$ search (red) \cite{ATLAS_Inv}, and monojet constraints (black) derived in \cite{Djouadi:2012zc} using the CMS monojet search \cite{Chatrchyan:2012me}. }
\label{Fig:fit_inv}
\end{center}
\vspace*{-3mm}
\end{figure}

 If the Higgs couplings to the SM matter take the SM values, then  the invisible width leads to a universal  reduction of the rates in all visible channels. 
This possibility is already quite constrained, given that we do see the Higgs produced with roughly the  SM rate.   
The left panel of  Fig.~\ref{Fig:fit_inv} shows $\Delta \chi^2$  as a function of the invisible branching fraction: ${\rm Br}_{\rm inv} >22$\% is excluded at  $95\%$ CL.\footnote{Simply combining the overall Higgs signal strengths quoted by  ATLAS  $\mu =1.30\pm 0.20$ \cite{ATLAS_coup}, CMS $\mu = 0.80 \pm 0.14$ \cite{CMSnew5}, and Tevatron $\mu = 1.44^{+0.59}_{-0.56}$ \cite{Aaltonen:2013kxa} one obtains the bound ${\rm Br}_{\rm inv} >23.6\%$ at  $95\%$ CL, fairly close to the result of our fit.} 
This bound can be relaxed if new physics modifies also the Higgs couplings such that the Higgs production cross section is enhanced. 
An example of such set-up is plotted in the right  panel of Fig.~\ref{Fig:fit_inv}, where we show the allowed region assuming the invisible Higgs  branching fraction and, simultaneously, a non-zero NLO coupling to gluons. Even in this  more general case ${\rm Br}_{\rm inv}$ larger than  $\sim 50\%$ is excluded at $95\%$ CL.
The indirect limits on the invisible width are in most cases much stronger than the direct ones from the ATLAS $Z+h \to {\rm inv.}$ search and from monojet searches.   

\section{Conclusions}
\label{sec:conclusions}

In this paper we updated the experimental constraints on the parameters of the Higgs effective Lagrangian. 
We combined the most recent LHC Higgs data in all available channels with the electroweak precision observables from SLC, LEP-1, LEP-2, and the Tevatron. 
Overall, the data are perfectly consistent with the 126 GeV particle discovered at the LHC  being the SM Higgs boson. 
A slight tension with the SM and a preference for negative Yukawa couplings found in previous Higgs fits goes away after including the latest CMS data in the $h \to \gamma \gamma$ channel.  
The leading order Higgs couplings to the SM matter in \eq{eq:lp2}  are well constrained, especially the coupling $c_V$ to  $W$ and $Z$ bosons which is constrained by a combination of Higgs and EW data to be within 10\%  of the SM value  at 95\% CL.   
The corollary is that the new particle is a Higgs boson:  it couples to the mass of the $W$ and $Z$ bosons, therefore it plays a role in electroweak symmetry breaking. 
This statement is independent of the size of possible higher-order Higgs couplings to W and Z (that play no role in electroweak symmetry breaking). 
 Higher-order  (2-derivative) couplings in the effective Lagrangian \eq{eq:lp4} are also well constrained by a combination of Higgs data and electroweak precision tests. 
 The couplings $c_{\gamma \gamma}$ and $c_{gg}$ to photons and gluons are constrained by Higgs data at the level of $10^{-3}$ (although some degeneracy between $c_{gg}$ and$c_t$ persists for the time being, lacking stringent constraints on the ttH production mode).   
 The coupling $c_{Z \gamma}$ is less  constrained, at the level of few$\times 10^{-2}$, by the combination of Higgs and electroweak data. 
 Other 2-derivative Higgs  couplings to gauge bosons must either  vanish or be tightly correlated with $c_{\gamma \gamma}$ $c_{Z \gamma}$,  barring large fine-tuning of electroweak precision observables.

\vspace{1cm}

\section*{Acknowledgements}

We thank E.~Kuflik, M.~Montull, A.~Pomarol and R.~Rattazzi for enlightening discussions. 
We are grateful to R.~Kogler from the Gfitter group and to W.~P{\l}aczek of BHWIDE for providing details about their programs.
FR acknowledges support from the Swiss National Science Foundation, under the Ambizione grant PZ00P2 136932 and thanks IFAE, Barcelona, for hospitality during completion of this work.  The work of AU is supported by the ERC Advanced Grant n$^{\circ}$ $267985$, ``Electroweak Symmetry Breaking, Flavour and Dark Matter: One Solution for Three Mysteries" (DaMeSyFla).

\appendix
\section{SILSH - Strongly Interacting Light Singlet Higgs}
\label{app:PMS}

In this section we present the Lagrangian,  parametrizing all relevant interactions between a scalar gauge singlet $h$ and a pair of gauge bosons,  in a basis where gauge invariance is manifest (but non-linearly realized). We then show how this relates to the phenomenological Lagrangian of Eqs.~(\ref{eq:lp2},\ref{eq:lp4}) and to the conditions \eq{eq:conditions} and classify the operators according to their expected size in a wide class of theories, in the spirit of Ref.~\cite{Giudice:2007fh}. While the latter discusses a strongly interacting light Higgs doublet (SILH), their arguments, which we follow closely in what follows, are useful also in the weakly coupled regime \cite{Elias-Miro:2013gya}.

The goldstone boson matrix $U\equiv \exp(i \varphi_j \tau^j/v)$ has well defined transformation properties under $SU(2)_L\times U(1)_Y$, with $\mathcal{G}_L=\exp(i\alpha_j\tau^j/2)$ and $\mathcal{G}_R=\exp(i\alpha_Y \tau^3/2)$,
\begin{equation}
U\to U^{\prime}=\mathcal{G}_L U \mathcal{G}_R^{\dag},
\end{equation}
and is the building-block for the Lagrangian with broken (non-linearly realized)  EW symmetry. Furthermore,
\begin{equation}
V_{\mu}\equiv(D_{\mu}U)U^{\dag},\quad \hat{W}_{\mu\nu}\equiv\partial_{\mu}\hat{W}_{\nu}-\partial_{\nu}\hat{W}_{\mu}-g[\hat{W}_{\mu},\hat{W}_{\nu}],\quad \textrm{and}\quad T=U\tau^3 U^{\dag},
\end{equation}
transform as adjoints of $SU(2)_L$ and singlets of $U(1)_Y$, where 
$D_{\mu}U=\partial_{\mu}U - g_L\hat{W}_{\mu}U+g_YU\hat{B}_{\mu}$ and $\hat{W}_{\mu}\equiv-\frac{i}{2}W_{\mu}^a\tau^a$, $\hat{B}_{\mu}\equiv-\frac{i}{2}B_{\mu}\tau^3$. The most general Lagrangian describing gauge bosons and $h$ can be written as an expansion in \footnote{Since $h$ is a singlet, interactions involving $h$ always imply coupling with a SM singlet operator involving gauge bosons. These have been classified in Ref.~\cite{Appelquist:1980vg} (Ref.~\cite{Longhitano:1980tm}) for the custodial preserving (breaking) case; see also Ref.~\cite{Alonso:2012px}.}
\begin{equation}
T,\,\, \frac{g_h h}{\Lambda},\,\, \frac{V_{\mu}}{\Lambda},\,\,\frac{\mathcal{D}_{\mu}}{\Lambda},\,\, \frac{g_L \hat{W}_{\mu\nu} }{\Lambda^2},\,\, \frac{g_Y B_{\mu\nu} }{\Lambda^2},
\end{equation}
where $B_{\mu\nu}$ is the usual $U(1)_Y$ field-strength tensor, $\mathcal{D}_{\mu}$ the covariant derivative and $g_h$ is the coupling of $h$ to the sector responsible to generate the corresponding term in the Lagrangian. In composite Higgs models, where $h$ arises as a resonance from a strong sector whose dynamics breaks the EW symmetry, we expect $g_h v \approx \Lambda $ (for simplicity we shall assume $g_h v/\Lambda=1$ in what follows, while we remark that the most general case can be obtained by accompanying each  insertion of $h$ with a factor $g_h v /\Lambda $).

We are interested in operators up to dimension 5, involving two gauge bosons and one scalar $h$, as these affect both EW precision tests and the $h$ decay widths.  At leading order in the derivative expansion, the only terms are
\begin{equation}\label{Ops1}
\mathcal{O}_{1}= -h \frac{v}{2}{\rm Tr}(V_{\mu}V^{\mu}), \quad \mathcal{O}_{1}^{\,\!\not\!\, C}=  h\frac{v}{2}{\rm Tr}(TV_{\mu}){\rm Tr}(TV^{\mu})
\end{equation}
and contribute as $c_{V}=c_1$ in \eq{eq:lp2}, where we denote in general by $c_i$ the coefficient of operator $\mathcal{O}_i$ in the Lagrangian. $\mathcal{O}_{1}^{\,\!\not\!\, C}$ breaks custodial symmetry even in the $g_Y\to0$ limit  and is absent in \eq{eq:lp2}. It is instructive to compare \eq{Ops1} with the analogous results of Ref.~\cite{Giudice:2007fh}, which makes the additional assumption that $h$ is part of an $SU(2)_L$ doublet that breaks the EW symmetry: we find the contribution to the above operators to be $c_1=c_H (v^2/f^2)$ and $c_{1}^{\,\!\not\!\, C}=c_T (v^2/f^2)$, where $c_{H,T}$ are the coefficients of the operators $\mathcal{O}_H=(\partial_\mu (H^\dagger H))^2/(2f^2)$ and $\mathcal{O}_T=(H^\dagger \overleftrightarrow {D^\mu} H)^2/(2f^2)$, with $f\equiv\Lambda/g_h$. 
Notice that, although the two contributions coincide in the limit $f\to v$, in realistic theories one finds $f\gg v$.

At the next order in the derivative expansion we have several contributions, which we classify accordingly to their expected size.
The operators\footnote{Other combinations can be eliminated using the identity
\begin{equation}
\mathcal{D}_{\mu}V_{\nu}-\mathcal{D}_{\nu}V_{\mu}
= -g_L\hat{W}_{\mu\nu} -\frac{ig_Y}{2}B_{\mu\nu}T+[V_{\mu},V_{\nu}]~.
\end{equation}
In particular ${\rm Tr}[T(\mathcal{D}_{\mu}\hat{W}^{\mu\nu})V_{\nu}]$ is a linear combination of $\mathcal{O}_{2}$, $\mathcal{O}_{2}^{\,\!\not\!\, C}$ and terms with more than two gauge bosons.}
\begin{eqnarray}\label{OpsTree}
\mathcal{O}_{2}&=&- g_L\frac{ v^2}{\Lambda^2}\frac{h}{v}{\rm Tr}[(\mathcal{D}_{\mu}\hat{W}^{\mu\nu})V_{\nu}],\quad
\mathcal{O}_{3}= g_Y \frac{ v^2}{\Lambda^2}\frac{h}{v}(\partial_{\mu}B^{\mu\nu}){\rm Tr}(TV_{\nu}), \nonumber\\
 \mathcal{O}_{2}^{\,\!\not\!\, C}&=&- g_L\frac{ v^2}{\Lambda^2}\frac{h}{v}\,{\rm Tr}[T(\mathcal{D}_{\mu}\hat{W}^{\mu\nu})]{\rm Tr}(T V_{\nu})
\end{eqnarray}
can arise at tree level mediated by a vector with mass $\Lambda$ and the appropriate quantum numbers \cite{Low:2009di}. They generate structures like $(h/v)Z_{\nu}(\partial_{\mu}Z^{\mu\nu})$, $(h/v)Z_{\nu}(\partial_{\mu}A^{\mu\nu})$, $(h/v)W^{\pm}_{\nu}(\partial_{\mu}W^{\mp,\mu\nu})$, already neglected in  \eq{eq:lp4} based on phenomenological arguments. Indeed, the operator $\mathcal{O}_{2}$ induces quadratic divergences in the S-parameter, while the operators  $\mathcal{O}_{2}^{\,\!\not\!\, C}$ and $\mathcal{O}_{3}$ induce power divergences in the T-parameter. Notice, furthermore, that  $\mathcal{O}_{3}$ does not formally break custodial invariance ($\mathcal{O}_{2}^{\,\!\not\!\, C}$ does), as it vanishes in the limit $g_Y\to 0$ and his smallness cannot be directly attributed to custodial symmetry. So, from a phenomenological point of view, all these operators are expected to be vanishingly small, as we assume in \eq{eq:lp4}. If $h$ comes from a Higgs doublet, we find the following contributions to the operators of \eq{OpsTree}: $c_2=c_W$ and $c_3=c_B$, where $c_{W,B}$ are the coefficients of
$\mathcal{O}_W\equiv(ig_L/(2\Lambda^2))\left( H^\dagger  \sigma^i \overleftrightarrow {D^\mu} H \right )( D^\nu  L_{\mu \nu})^i$ and 
$\mathcal{O}_B\equiv (ig_Y/(2\Lambda^2))\left( H^\dagger  \overleftrightarrow {D^\mu} H \right )( \partial^\nu  B_{\mu \nu})$ defined in Ref.~\cite{Giudice:2007fh}.

The operators
\begin{eqnarray}
\mathcal{O}_{4}&=& \frac{g_L^2 }{16 \pi^2}\frac{h}{v}{\rm Tr}(\hat{W}_{\mu\nu}\hat{W}^{\mu\nu}),\quad \mathcal{O}_{5}= \frac{g_Y^{2} }{16 \pi^2}\frac{h}{v}{\rm Tr}(\hat{B}_{\mu\nu}\hat{B}^{\mu\nu}),\quad
\mathcal{O}_{6}=  \frac{g_Lg_Y }{16 \pi^2}\frac{h}{v} B_{\mu\nu}{\rm Tr}(T\hat{W}^{\mu\nu})\nonumber\\
\mathcal{O}_{4}^{\,\!\not\!\, C}&=& \frac{g^2 }{16 \pi^2}\frac{h}{v}{\rm Tr}(T\hat{W}_{\mu\nu}){\rm Tr}(T\hat{W}^{\mu\nu})\label{Ops:Loop}
\end{eqnarray}
are instead expected to arise at the loop level in minimally coupled theories (i.e. theories where gauge bosons couple only through covariant derivatives \cite{Giudice:2007fh,Low:2009di,Elias-Miro:2013gya}). These appear in the phenomenological Lagrangian  \eq{eq:lp2} as 
$c_{\gamma\gamma}=e^2(2c_4-4c_5)/16\pi^2$ and 
$c_{Z\gamma}= e^2(c_4 g_L/g_Y +c_5g_Y/g_L)/16\pi^2$, while the custodial protecting conditions \eq{eq:conditions} correspond to the vanishing of  $\mathcal{O}_{6}$ and $\mathcal{O}_{4}^{\,\!\not\!\, C}$ (notice that the gauge-invariant operators reproduced by the sum rules \eq{eq:conditions}  are those that preserve the custodial symmetry even in the $g_Y\neq 0$ limit). A Higgs doublet would contribute to these operators as $c_4=-c_{HW}(v^2/f^2)$, $c_5=(c_\gamma-c_{HB}/4)(v^2/f^2)$ and $c_6=c_{HB}v^2/(2f^2)$ where $c_{HW,HB,\gamma}$ are the coefficients of the operators  \cite{Giudice:2007fh}
\begin{eqnarray}
\mathcal{O}_{HW}&\equiv \frac{i g_L}{16\pi^2f^2}
(D^\mu H)^\dagger \sigma^i(D^\nu H)L_{\mu \nu}^i,\quad 
\mathcal{O}_{HB}\equiv\frac{ig_Y}{16\pi^2f^2}
(D^\mu H)^\dagger (D^\nu H)B_{\mu \nu},\nonumber\\
& \mathcal{O}_{\gamma}\equiv\frac{{g_Y}^2}{16\pi^2f^2}H^\dagger H B_{\mu\nu}B^{\mu\nu}.
 \end{eqnarray}
Notice that, if the Higgs doublet arises as a pseudo Nambu-Goldstone boson of a strongly interacting sector \cite{Contino:2010rs}, then $\mathcal{O}_\gamma$ is further suppressed by $\sim g^2/g_h^2$, implying a smaller $c_{\gamma\gamma}$ in \eq{eq:lp4}.

Finally the operators 
\begin{eqnarray}\label{ops:lastones}
\mathcal{O}_{7}&=&\frac{v^2 }{\Lambda^2}\frac{h}{v}{\rm Tr}[(\mathcal{D}_{\mu}V_{\nu})(\mathcal{D}^{\mu}V^{\nu})]~,\quad
\mathcal{O}_{8}=\frac{v^2 }{\Lambda^2}\frac{h}{v}{\rm Tr}[(\mathcal{D}^{\nu}\mathcal{D}_{\mu}V_{\nu})V^{\mu})]~,\\
\mathcal{O}_{7}^{\,\!\not\!\, C}&=&\frac{v^2 }{\Lambda^2}\frac{h}{v}{\rm Tr}[(T\mathcal{D}_{\mu}V_{\nu})(T\mathcal{D}^{\mu}V^{\nu})]~,\quad
\mathcal{O}_{8}^{\,\!\not\!\, C}=\frac{v^2 }{\Lambda^2}\frac{h}{v}{\rm Tr}[T(\mathcal{D}^{\nu}\mathcal{D}_{\mu}V_{\nu})]{\rm Tr}(TV^{\mu})\,,\nonumber
\end{eqnarray}
complete our list\footnote{The operator $(h/v){\rm Tr}[T(\mathcal{D}_{\mu}V^{\mu})]{\rm Tr}[T(\mathcal{D}_{\nu}V^{\nu})]$ and his custodial breaking analogue, have no influence on EW precision parameters, nor on the Higgs decay widths.}. These terms are in principle expected to be larger than the ones in \eq{Ops:Loop}, coinciding only in the limit $g_h\sim \Lambda/v \to 4\pi$. Nevertheless it is interesting to note that, if $h$ is part of an $SU(2)_L$ doublet and if we assume that the sector responsible for generating these couplings only involves particles with spin $\leq 1$, then contribution to these terms do not arise at dimension 6, but rather at dimension 8 and are therefore  suppressed by further powers of $g_h v/\Lambda=v/f$ which is are generally small \cite{Giudice:2007fh}. This is also what is expected from a phenomenological point of view (they generate structures like $(h/v)\partial_\mu Z_\nu\partial_\nu Z_\mu$), since they contribute to the EW precision observables via quadratic divergences.

In summary, we have listed  four classes of interactions between two gauge bosons and a singlet scalar: the leading ones Eq.~(\ref{Ops1}) in the derivative expansion  are expected to have the bigger effect; among the subleading ones, Eq.~(\ref{OpsTree}) can arise at tree-level, while Eq.~(\ref{Ops:Loop}) arise typically at loop-level and Eq.~(\ref{ops:lastones}) arise only at dimension 8 in models where $h$ is part of a doublet. 

In our phenomenological analysis we have kept only the leading interactions and the ones that arise at loop-level, as the ones of Eqs.~(\ref{OpsTree},\ref{ops:lastones}) induce power-divergences in the EW parameters.
 At first sight it is unpleasant that the operators that are supposed to be larger from an Naive Dimensional Analysis point of view (that is, $\mathcal{O}_{2,3}$), are neglected in favor of $\mathcal{O}_{4,5}$, which are expected to arise only at the loop level. This however could have been expected if we take into account the fact that, since $h$ is a singlet, the existence of $\mathcal{O}_{2,3}$ implies in principle the existence of operators  ${\rm Tr}[T(\mathcal{D}_{\mu}\hat{W}^{\mu\nu})V_{\nu}]$ and $\partial_{\mu}B^{\mu\nu}{\rm Tr}(TV_{\nu})$, which contribute at tree level to the EW precision parameters and are forced to be small. We are therefore assuming that the same dynamical mechanism that forbids tree-level effects, also accounts for the suppression of $h$-loop mediated effects.

\section{Oblique Parameters}
\label{app:PO}

The term {\em oblique corrections}  refers to modifications of the propagators of the electroweak gauge bosons. 
In many models beyond the SM the largest new contributions to physical observables enter via the oblique corrections  (for a more general approach to electroweak data, see  \cite{Han:2004az}). 
Let us define an expansion of the  2-point functions in powers of $p^2$, where $p$ is the 4-momentum flowing through the diagram:   
\beq
\cM(V_{1,\mu} \to V_{2,\nu}) =   \eta_{\mu\nu} \left ( \Pi_{V_1 V_2}^{(0)} +  \Pi_{V_1V_2}^{(2)} p^2 +  \Pi_{V_1V_2}^{(4)} p^4 +   \dots \right )  
+ p_{\mu}p_{\nu}  \left ( \dots \right)
\eeq
where $V_i = 1,2,3,B$ label the $SU(2)_L \times U(1)_Y$ gauge bosons on the external lines. 
Two-point functions are not directly  measurable, but certain combinations of  $\Pi_{V_1 V_2}$ affect measurable quantities. 
Up to order $p^2$ the physical combinations are the 3 Peskin--Takeuchi oblique parameters \cite{Peskin:1991sw} defined  as
\beq
\label{eq:st_ptp} 
\alpha S =   
- 4 {g_L g_Y \over g_L^2 + g_Y^2} \delta \Pi_{3 B}^{(2)},
\quad 
\alpha T  =  
{\delta \Pi_{11}^{(0)} - \delta \Pi_{33}^{(0)}  \over m_W^2}, 
\quad 
\alpha U =  
{4 g_Y^2 \over g_L^2 + g_Y^2}   \left (\delta \Pi_{11}^{(2)} - \delta \Pi_{33}^{(2)} \right ).  
\eeq 
where $\delta \Pi$ denotes a shift of the corresponding 2-point function from the SM value, and the fine-structure constant $\alpha \approx 1/137$ is used for normalization. 
At order $p^4$ one can define further oblique parameters \cite{Barbieri:2004qk}.\footnote{Compared to Ref.~\cite{Barbieri:2004qk} we rescaled these parameters by $\alpha$ so as to put them numerically on equal footing with $S$, $T$, $U$. }  
\beq
\alpha V =    m_W^2   \left (\delta \Pi_{11}^{(4)} - \delta \Pi_{33}^{(4)} \right ), 
\quad 
\alpha W  =   - m_W^2   \delta \Pi_{33}^{(4)}, 
\quad 
\alpha X  =   -  m_W^2   \delta \Pi_{3B}^{(4)}, 
\quad 
\alpha Y  =   -  m_W^2  \delta  \Pi_{BB}^{(4)}.  
\eeq
LEP-1 and SLC measurements on the Z resonance constrain two linear combinations of the 2 point-functions, and the W mass measurement constrains another. 
Thus, by itself, the above measurement can constrain only three obliques parameters, e.g.  $S$, $T$ and $U$, which explains the origin of the Peskin--Takeuchi parametrization.  
Adding the constraints from off-Z-pole measurements in LEP-2 allows one to put meaningful limits on $V$, $W$, $X$, and $Y$ as well.    
Out of these seven oblique parameters four are singled out because, assuming the  Higgs field is an $SU(2)_L$ doublet $H$, they correspond to dimension-6 operators beyond the SM.
The mapping is 
\beq
\cl_{\rm dim-6} =  
{\alpha S (g_L^2 + g_Y^2)  \over 4 v^2 g_L g_Y } (H^\dagger \sigma^a H) W_{\mu\nu}^a B_{\mu \nu} 
- {2 \alpha T  \over v^2}  |H^\dagger D_\mu H|^2  
-  {\alpha W \over 4 m_W^2} (D_\rho W_{\mu \nu}^a)^2   - {\alpha Y \over 4 m_W^2} (\pa_\rho B_{\mu \nu})^2  
\eeq 
whereas $U$, $V$, $X$ correspond to operators of dimension 8 and higher. 
For this reason, typical new physics  models affect $S$, $T$, $W$, $Y$ in the first place.
Combining the EW observables listed in Table~\ref{default} for $U=V=X=0$ we obtain the following constraints: 
\beq
S = 0.05 \pm 0.18, 
\quad 
T = 0.09 \pm 0.09, 
\quad
W = -0.02 \pm 0.09, 
\quad  
Y = 0.00 \pm 0..21 \, .   
\eeq 
Moreover, setting also $W= Y= 0$ we find in $S = 0.05 \pm 0.09$, $T = 0.09 \pm 0.07$, in agreement with the Gfitter result  \cite{Baak:2012kk} (for a fit to $S$ and $T$ alone the LEP-2 constraints, not included by Gfitter, are not important).

\end{document}